\documentclass[onefignum,onetabnum]{siamart171218}

\usepackage[T1]{fontenc}
\usepackage[utf8]{luainputenc}
\usepackage{float}
\usepackage{amsfonts}
\usepackage{amsmath}
\usepackage{amssymb}
\usepackage{amsopn}
\usepackage{graphicx}
\usepackage{color}
\usepackage{subfig}
\usepackage{algorithm}
\usepackage[page]{appendix}
\usepackage{hyperref}
\usepackage{graphicx}

\newsiamremark{remark}{Remark}

\renewcommand{\theequation}{\thesection.\arabic{equation}}
\numberwithin{equation}{section}

\begin{document}

\title{Recovering Hidden Components in Multimodal Data with Composite Diffusion Operators}
\author{Tal Shnitzer\thanks{Andrew and Erna Viterbi Faculty of Electrical Engineering, Technion -- Israel Institute of Technology, Technion City, Haifa, Israel 3200000, (\email{shnitzer@campus.technion.ac.il}, \email{ronen@ee.technion.ac.il}).}
\and Mirela Ben-Chen\thanks{Department of Computer Science, Technion -- Israel Institute of Technology, Technion City, Haifa, Israel 3200000, (\email{mirela@cs.technion.ac.il}).}
\and Leonidas Guibas\thanks{Department of Computer Science, Stanford University, Stanford, CA 94305, USA, (\email{guibas@cs.stanford.edu}).}
\and Ronen Talmon\footnotemark[1]
\and Hau-Tieng Wu\thanks{Department of Mathematics and Department of Statistical Science, Duke University, Durham, NC 27708, USA, (\email{hauwu@math.duke.edu})}}

\maketitle

\begin{abstract}
{Finding appropriate low dimensional representations of high-dimensional multi-modal data can be challenging, since each modality embodies unique deformations and interferences. 
In this paper, we address the problem using manifold learning, where the data from each modality is assumed to lie on some manifold. In this context, the goal is to characterize the relations between the different modalities by studying their underlying manifolds.
We propose two new diffusion operators that allow to isolate, enhance and attenuate the hidden components of multi-modal data in a data-driven manner. Based on these new operators, efficient low-dimensional representations can be constructed for such data, which characterize the common structures and the differences between the manifolds underlying the different modalities. 
The capabilities of the proposed operators are demonstrated on 3D shapes and on a fetal heart rate monitoring application.
}
\end{abstract}

\section{Introduction}
Recent technological progress leads to highly heterogeneous datasets, consisting of multimodal samples acquired by a multitude of sensors. Current research is plagued by the problem of finding the ``appropriate'', often low dimensional, representation for such high-dimensional multimodal data. Indeed, obtaining meaningful representations from multimodal data is truly challenging, since such data comprise many latent sources of variability, each source embodies unique and possibly redundant information; while some of these sources are important, some are completely superfluous. This naturally leads to questions such as how to discover and isolate the different sources, how to identify and extract the relevant information, and how to merge data from different modalities. 

Various studies have addressed multimodal data analysis problems \cite{lahat2015multimodal}. A few examples include the classical Canonical Correlation Analysis (CCA) \cite{hotelling1936relations}, which recovers highly correlated linear projections from two datasets, and recent CCA extensions which involve kernels to address nonlinearities \cite{lai2000kernel,andrew2013deep,michaeli2016nonparametric}. 
Methods relying on kernels are of particular interest in the context of the present work. For example, methods for spectral clustering of multimodal data based on kernel manipulation are presented in \cite{zhou2007spectral,kumar2011co,de2005spectral}. 
In \cite{zhou2007spectral}, spectral clustering is performed on the multimodal data by solving the generalized eigenvalue problem of a new matrix, constructed based on a mixture of random walks defined on multiple graphs, each representing a different view.
In \cite{kumar2011co}, multimodal spectral clustering is learned by iteratively clustering each view separately and then modifying the graph structures accordingly. 
Another work \cite{de2005spectral}, combines affinity matrices of two graphs, representing two different views, by constructing a larger symmetric affinity matrix, which is based on their multiplication.
Other related work includes (i) the construction of a joint manifold by concatenating samples from several sensors, each represented by a separate manifold \cite{davenport2010joint}, (ii) metric fusion obtained by combining similarity measures through kernel multiplication \cite{wang2012unsupervised}, and (iii) a new representation of multiview data learned by jointly diagonalizing Laplacians of different views \cite{eynard2015multimodal}. In addition, \cite{coifman2014diffusion} presents a method for mapping low dimensional graph Laplacian representations of different views (or times) into a common latent space, allowing for the analysis of multimodal data in a low dimensional intrinsic space.

Our specific focus here is on a manifold learning approach. Consider a single high-dimensional dataset assumed to live on a single manifold. Analyzing this dataset with typical manifold learning methods, such as laplacian eigenmaps \cite{belkin2003laplacian} or diffusion maps \cite{Coifman2006}, simplifies to computing a kernel based on an affinity suitable for the dataset at hand. Then, by employing spectral analysis, the data are embedded in a new Euclidean space that captures their underlying manifold structure. The natural question then arises -- are the required mathematical properties for spectral analysis transferable to settings comprising several datasets? If this could be achieved, the data analysis procedure could be naturally extended to analyze multiple datasets, deforming the intrinsic space in different ways. 

Apparently, manifold learning techniques almost exclusively address only a single manifold structure. In a recent work \cite{lederman2015learning}, a data-driven method for recovering the common latent variable underlying multiple, multimodal sensor data based on alternating products of diffusion operators was presented. 
This work was later extended in \cite{talmon2016latent}, showing that the alternating products of diffusion operators recovers a common manifold structure. 
In addition, as proven in \cite{talmon2016latent,lederman2015learning}, it ignores the components specific to each modality. 
However, the product of diffusion operators does not necessarily have a real spectrum. 
Other recent work \cite{froyland2015dynamic,marshall2017time} propose to analyze dynamical systems based on products of diffusion operators, in a manner related to \cite{lederman2015learning}. There, data from each time frame is modeled as samples from a manifold with a time-evolving metric, and by revealing the common latent variables of several time frames, they recover coherent sets (in \cite{froyland2015dynamic}) or a representation of the common latent manifold in time (in \cite{marshall2017time}).

In this paper, we propose new diffusion operators defined on data arising from multiple sensors, allowing for a nonlinear efficient data-driven way to isolate, enhance and attenuate various hidden components. More concretely, we propose two operators that reveal the common structures and the differences between manifolds. We show that these two operators have a meaningful spectral decomposition, which we leverage to construct an efficient low-dimensional representation.

The capabilities of the presented operators in extracting hidden components are demonstrated in simulations and on a real-world application to fetal heart rate monitoring. Fetal heart rate monitoring is widely-used for the assessment of the fetus' health both during pregnancy and during delivery. The most accurate method, relying on the placement of electrodes on the fetus' scalp, is invasive, and therefore, carries many risks. Consequently, non-invasive measurements are usually carried out by placing electrodes on the abdomen of the mother (see a comprehensive review in \cite{Sameni2010}). Naturally, the measured signal contains, in addition to the fetal electrocardiogram (ECG), the maternal ECG, masking the desired information. In order to suppress the maternal ECG and to extract the fetal ECG, common practice is to use another (reference) electrode, placed on the mother's thorax, for the purpose of measuring only the maternal ECG. Then, the relation between the measured abdomen and thorax signals is extracted, using, for example, the adaptive least mean squares (LMS) algorithm \cite{widrow1976stationary}. In this work, we detect the fetal ECG from two abdomen signals, which is considered a challenging problem that does not have a definitive solution to date. We show that the proposed operators discover the relations between the signals acquired with multiple sensors in a data-driven manner, revealing their hidden components.

\section{Problem formulation\label{sec:formulation}}
Consider two diffeomorphic manifolds, $\mathcal{M}^{(1)}$ and $\mathcal{M}^{(2)}$, with a diffeomorphism $\phi:\mathcal{M}^{(1)}\mapsto\mathcal{M}^{(2)}$, where each manifold $\mathcal{M}^{(\ell)}$ is a compact Riemannian manifold without a boundary of dimension $d$ with a metric $g^{(\ell)}$.
In this work, we will distinguish between the following two structures: 
\begin{align}
\Omega_{\alpha} &= \left\{ x\in\mathcal{M}^{(1)}: \ \nabla\phi |_x = \alpha\mathrm{I} \right\} \subset \mathcal{M}^{(1)} \label{eq:defOmega}\\ 
\Omega^c_{\alpha} &= \mathcal{M}^{(1)} \backslash \mathring{\Omega}_{\alpha},\label{eq:defOmegaC}
\end{align}
where $\alpha>0$ is a scaling factor, $\mathrm{I}$ denotes a $d\times d$ identity matrix, $\nabla\phi |_x$ is represented by a pair of properly chosen orthonormal bases at $T_x\mathcal{M}^{(1)}$ and $T_{\phi(x)}\mathcal{M}^{(2)}$, and $\mathring{\Omega}_{\alpha}$ denotes the maximal open subset of the closed set $\Omega_{\alpha}$. Therefore, $\Omega_{\alpha}$ denotes all structures which are similar, up to a scaling $\alpha>0$, in the two manifolds, $\mathcal{M}^{(1)}$ and $\mathcal{M}^{(2)}$.

Our goal is to identify and isolate $\Omega_{\alpha}$ and $\Omega^c_{\alpha}$ in a data-driven manner, given pairs of observation samples $(x,y)$, such that $x \in \mathcal{M}^{(1)}$, $y \in \mathcal{M}^{(2)}$, and $y=\phi(x)$.
We will show in the sequel that the two structures  $\Omega_{\alpha}$ and $\Omega^c_{\alpha}$ have great importance in data analysis problems.

\begin{figure}
\centering
\subfloat[]{\includegraphics[width=0.4\textwidth]{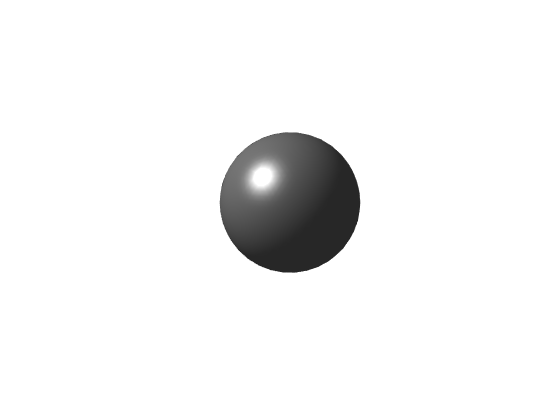}}
\subfloat[]{\includegraphics[width=0.4\textwidth]{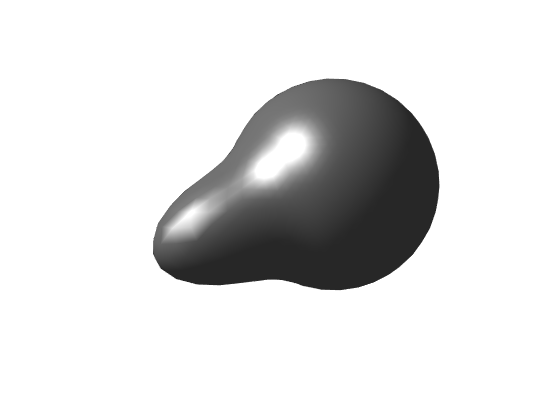}}
\caption{Two diffeomorphic geometric shapes. A sphere (a) and a scaled sphere with a deformation (a ``bump'') (b). \label{fig:Spheres}}
 \end{figure}

For example, consider the two geometric shapes presented in Figure \ref{fig:Spheres}. Figure \ref{fig:Spheres}(a) depicts a 2-sphere and Figure \ref{fig:Spheres}(b) depicts a scaled and deformed sphere, i.e., a scaled sphere with a ``bump''. Denote these two shapes by $\mathcal{M}^{(1)}$ and $\mathcal{M}^{(2)}$ respectively. The deformation and scaling of $\mathcal{M}^{(2)}$ can be represented by a diffeomorphism between the two shapes $\phi:\mathcal{M}^{(1)}\to\mathcal{M}^{(2)}$. In this example, by definition, the undeformed sphere structure (up to scaling) is represented by $\Omega_{\alpha}$ and the ``bump''  is represented by $\Omega^c_{\alpha}$. Therefore, given the two shapes, our goal is to recover a separate representation for $\Omega_{\alpha}$ and $\Omega^c_{\alpha}$.
 
This problem formulation, describing common structures of two manifolds, i.e. $\Omega_{\alpha}$, can be seen as analogous to recent work \cite{froyland2015dynamic,froyland2017dynamic}. There, a framework for recovering coherent sets in dynamical systems is proposed, where each time instance is represented by some underlying manifold, $\mathcal{M}$, and the system dynamics are represented by a diffeomorphism, $\phi$. Since coherent sets represent system behavior that changes slowly in time, they can be described by the common structures, i.e. $\Omega_{\alpha}$ in our formulation.

\section{Diffusion operators for multimodal data\label{sec:theory}}
In this section, we present the derivation of the proposed operators, starting from a single manifold setting in Subsection \ref{sub:TheoSingle}, similarly to \cite{Coifman2006}. In Subsection \ref{sub:theoAd}, we present an extension to two manifolds, as a variant of \cite{talmon2016latent,lederman2015learning}, and finally, in Subsection \ref{sub:theoNew}, we present the proposed new operators for revealing the common and difference structures of two manifolds.

\subsection{Preliminaries -- single manifold setting\label{sub:TheoSingle}}
Define the following symmetric kernel for a manifold $\mathcal{M}$, based on its distance function, denoted by $d_{g}$, corresponding to the metric $g$ on $\mathcal{M}$ by
\begin{equation}
k_{\epsilon}\left(x,x'\right)=\exp\left(-\frac{d_{g}\left(x,x'\right)^2}{\epsilon^2}\right),\label{eq:dmkern}
\end{equation}
where $x,x'\in\mathcal{M}$.
The kernel is then normalized by 
\begin{equation}
p_{\epsilon}\left(x,x'\right)=\frac{k_{\epsilon}\left(x,x'\right)}{d_{\epsilon}\left(x\right)},
\end{equation}
where $d_{\epsilon}\left(x\right)=\int k_{\epsilon}\left(x,x'\right)\mu\left(x'\right)dV(x')$, $V$ is the volume measure induced by $g$, and $\mu\left(x'\right)$ is the density function of the points on $\mathcal{M}$. Similarly, define the following normalized kernel by
\begin{equation}
q_{\epsilon}\left(x,x'\right)=\frac{k_{\epsilon}\left(x,x'\right)}{d_{\epsilon}\left(x'\right)}.
\end{equation}

Based on $q_{\epsilon}\left(x,x'\right)$ and $p_{\epsilon}\left(x,x'\right)$, we define the following \textit{``backward''} and \textit{``forward''} diffusion operators
\begin{align}
P_{\epsilon}f\left(x\right) &= \int p_{\epsilon}\left(x,x'\right)f\left(x'\right)\mu\left(x'\right)dV(x') \\
Q_{\epsilon}f\left(x\right) &= \int q_{\epsilon}\left(x,x'\right)f\left(x'\right)\mu\left(x'\right)dV(x')
\end{align}
for any $f \in C^{\infty}(\mathcal{M})$. 

\begin{proposition}\label{prop:PQ}
Suppose $\mu,f\in C^4(\mathcal{M})$, where $\mathcal{M}$ is a smooth Riemannian manifold with a metric $g$. The asymptotic expansion of the operators $P_\epsilon$ and $Q_\epsilon$, when $\epsilon$ is sufficiently small, is given by
\begin{align}
P_{\epsilon}f\left(x\right) &= f(x) - \epsilon^2 \left(\Delta f + \frac{2\nabla f\cdot\nabla\mu}{\mu} \right) (x) + O(\epsilon^4) \\
Q_{\epsilon}f\left(x\right) &= f(x) - \epsilon^2\left(\Delta f - \frac{f\Delta\mu}{\mu}\right) (x) + O(\epsilon^4),
\end{align}
where $\nabla$ denotes the covariant derivative on the manifold, $\mathcal{M}$, and $\Delta$ denotes the Laplace-Beltrami operator.
\end{proposition}
This derivation of the backward operator, $P_\epsilon$, is shown in \cite{Coifman2006} and the derivation of the forward operator, $Q_\epsilon$, is shown in Appendix \ref{app:propPQ}.

The operator $Q_\epsilon$ is the \textit{forward} operator, similarly defined in \cite{nadler2006diffusion}, which can be interpreted as an operator that propagates probability density functions on the manifold in time. Operator $P_\epsilon$ is the \textit{backward} operator, which can be interpreted as propagating averages of functions on the manifold in time. These two operators are adjoint under the inner product with $\mu$ \cite{nadler2006diffusion,Coifman2006}.

\begin{sloppypar}
From the spectral decomposition of the operators $P_\epsilon$ and $Q_\epsilon$, a new low-dimensional representation for $\mathcal{M}$ is typically obtained, which approximates the diffusion distance between data points, $x,x'\in\mathcal{M}$, as described in \cite{Coifman2006}. The backward operator $P_\epsilon$ was previously used in numerous applications to recover a meaningful representation of the data (e.g. \cite{talmon2013diffusion,talmon2013empirical,lafon2006data}).
\end{sloppypar}

\subsection{Modified alternating diffusion in a two manifold setting\label{sub:theoAd}}
Given two manifolds, denoted by $\mathcal{M}^{(1)}$ and $\mathcal{M}^{(2)}$, consider the following $C^{\infty}(\mathcal{M}^{(1)}) \rightarrow C^{\infty}(\mathcal{M}^{(1)})$ operators:
\begin{align}
G_{\epsilon_1,\epsilon_2} f(x) &= \phi^{*} P_{\epsilon_2}^{(2)} (\phi^{*})^{-1} Q_{\epsilon_1}^{(1)} f(x) \\
H_{\epsilon_1,\epsilon_2} f(x) &= P_{\epsilon_1}^{(1)} \phi^{*} Q_{\epsilon_2}^{(2)} (\phi^{*})^{-1} f(x)
\end{align}
for any function $f \in C^{\infty}(\mathcal{M}^{(1)})$, where $\epsilon_1,\epsilon_2>0$, $\phi^{*}:C^{\infty}(\mathcal{M}^{(2)})\rightarrow C^{\infty}(\mathcal{M}^{(1)})$ denotes the operator corresponding to the pullback from $\mathcal{M}^{(2)}$ to $\mathcal{M}^{(1)}$, i.e., $\left(\phi^*g\right)(x)=\left(g\right)\left(\phi(x)\right)$ for $x\in\mathcal{M}^{(1)}$, $g \in C^{\infty}(\mathcal{M}^{(2)})$, and $(\phi^{*})^{-1}$ denotes the pullback from $\mathcal{M}^{(1)}$ to $\mathcal{M}^{(2)}$, which inverts $\phi^*$. 

Note that for such a composition of operators, the interpretation of the forward operator as propagating probability density functions does not extend to the operators $G_\epsilon$ and $H_\epsilon$. 
In the following proposition, we present an analysis for the new operators, $G_{\epsilon_1,\epsilon_2}$ and $H_{\epsilon_1,\epsilon_2}$, which are the composition of $P_{\epsilon_\ell}^{(\ell)}$ and $Q_{\epsilon_\ell}^{(\ell)}$, $\ell=1,2$, based on their asymptotic expansions.

\begin{proposition}\label{prop:GH}
When $\epsilon_1,\epsilon_2>0$ are sufficiently small and $\mu^{(1)}$ is smooth enough, the asymptotic expansions of the operators $G_{\epsilon_1,\epsilon_2}$ and $H_{\epsilon_1,\epsilon_2}$ are given by
\begin{align}
G_{\epsilon_1,\epsilon_2} f(x) = & f(x)  - \epsilon_1^2\Delta ^{(1)}f(x) - \epsilon_2^2\phi^*\Delta^{(2)}(\phi^{*})^{-1}f(x)\label{eq:G1}\\
& - \epsilon_2^2\phi^*\frac{2\nabla^{(2)}(\phi^{*})^{-1}f\cdot\nabla^{(2)}\mu^{(2)}}{\mu^{(2)}}(x) + \epsilon_1^2\frac{f\Delta^{(1)}\mu^{(1)}} {\mu^{(1)}}(x) + O(\epsilon_1^4+\epsilon_2^4)\label{eq:G2}\\
H_{\epsilon_1,\epsilon_2} f(x) = & f(x) - \epsilon_1^2\Delta ^{(1)}f(x) - \epsilon_2^2\phi^*\Delta^{(2)}(\phi^{*})^{-1}f(x)\label{eq:H1}\\
& - \epsilon_1^2\frac{2\nabla^{(1)}f\cdot\nabla^{(1)}\mu^{(1)}}{\mu^{(1)}}(x) + \epsilon_2^2\phi^*\frac{\Delta^{(2)}\mu^{(2)}} {\mu^{(2)}}(x) + O(\epsilon_1^4+\epsilon_2^4).\label{eq:H2}
\end{align}
\end{proposition}
The derivations for both operators appear in Appendix \ref{app:propGH}.

Note that the asymptotic expansion of these operators can be described by a term which depends on the geometry, the Laplace-Beltrami operators $\Delta^{(1)}$ and $\Delta^{(2)}$ in both \eqref{eq:G1} and \eqref{eq:H1}, and a term which depends on both the geometry and the densities, $\mu^{(1)}$, $\mu^{(2)}$, in both \eqref{eq:G2} and \eqref{eq:H2}.

\begin{sloppypar}
In \cite{lederman2015learning,talmon2016latent}, alternating diffusion operators are defined in a related manner. In \cite{lederman2015learning}, the operator $\phi^*Q_{\epsilon_2}^{(2)}(\phi^{*})^{-1}Q_{\epsilon_1}^{(1)}$ was introduced, and in \cite{talmon2016latent} the operator $\phi^*P_{\epsilon_2}^{(2)}(\phi^{*})^{-1}P_{\epsilon_1}^{(1)}$ was studied. Both variants are compositions of two operators, each corresponding to a different manifold. It was shown there that these operators reveal the common structure of the two manifolds. Note that the alternating diffusion operators are different than the operators proposed here, due to the use of two backward or forward operators in alternating diffusion, instead of one backward and one forward operator, as proposed here. We will show that the modification considered here is not only semantic and it leads to a different asymptotic behavior than the one described in \cite{talmon2016latent}. 
The difference between the asymptotic expansions in \eqref{eq:G2} and \eqref{eq:H2}, and the corresponding asymptotic expansion of the alternating diffusion operator is described in detail in Appendix \ref{app:compareAD}. 
\end{sloppypar}

\subsection{Composite operators in a two manifold setting\label{sub:theoNew}}
The operators in Subsection \ref{sub:theoAd} and in \cite{talmon2016latent,lederman2015learning} suffer from several shortcomings. First, as presented in \cite{talmon2016latent}, the alternating diffusion operator highly depends on the order of the kernel multiplication (in a realistic discrete setting). Note that this is also true for operators $G_{\epsilon_1,\epsilon_2}$ and $H_{\epsilon_1,\epsilon_2}$, which depend on the kernel order even in the continuous setting, as portrayed by their asymptotic expansions.
Second, these operators are not self-adjoint nor normal (see Appendix \ref{app:compareAD}) and therefore, the spectral theorem does not hold. 
In this subsection, we address these problems and propose two new operators, $S_{\epsilon_1,\epsilon_2}$, which will be shown to reveal \emph{common structures}, and $A_{\epsilon_1,\epsilon_2}$, which will be shown to reveal \emph{differences}. 

Define
\begin{align}
S_{\epsilon_1,\epsilon_2} f(x) &= \frac{1}{2}\left(G_{\epsilon_1,\epsilon_2} f(x) + H_{\epsilon_1,\epsilon_2} f(x)\right) \\
A_{\epsilon_1,\epsilon_2} f(x) &= \frac{1}{2} \left(G_{\epsilon_1,\epsilon_2} f(x) - H_{\epsilon_1,\epsilon_2} f(x)\right) 
\end{align}

\begin{proposition}\label{prop:SA}
When $\epsilon_1,\epsilon_2>0$ are sufficiently small and $\mu^{(1)}$ is smooth enough, the asymptotic expansions of the operators $S_{\epsilon_1,\epsilon_2}$ and $A_{\epsilon_1,\epsilon_2}$ are given by 
\begin{align}
S_{\epsilon_1,\epsilon_2} f(x) = & f(x) - \epsilon_1^2\Delta^{(1)}f(x) -\epsilon_2^2\phi^*\Delta^{(2)}(\phi^{*})^{-1}f(x)\label{eq:s_exp1}\\
& - \frac{\epsilon_2^2}{2}\left(\phi^*\frac{2\nabla^{(2)}(\phi^{*})^{-1}f\cdot\nabla^{(2)}\mu^{(2)}}{\mu^{(2)}}(x) - f\phi^*\frac{\Delta^{(2)}\mu^{(2)}}{\mu^{(2)}}(x)\right)\\
& - \frac{\epsilon_1^2}{2}\left(\frac{2\nabla^{(1)}f\cdot\nabla^{(1)}\mu^{(1)}}{\mu^{(1)}}(x) - \frac{f\Delta^{(1)}\mu^{(1)}}{\mu^{(1)}}(x)\right) + O(\epsilon_1^4+\epsilon_2^4)\\
A_{\epsilon_1,\epsilon_2} f(x) = & \frac{\epsilon_1^2}{2}\left(\frac{2\nabla^{(1)}f\cdot\nabla^{(1)}\mu^{(1)}}{\mu^{(1)}}(x) + \frac{f\Delta^{(1)}\mu^{(1)}}{\mu^{(1)}}(x)\right)\label{eq:a_exp1}\\
& - \frac{\epsilon_2^2}{2}\left(\phi^*\frac{2\nabla^{(2)}(\phi^{*})^{-1}f\cdot\nabla^{(2)}\mu^{(2)}}{\mu^{(2)}}(x) + f\phi^*\frac{\Delta^{(2)}\mu^{(2)}} {\mu^{(2)}}(x)\right) + O(\epsilon_1^4+\epsilon_2^4)\label{eq:a_exp2}
\end{align}
The derivations for both operators appear in Appendix \ref{app:propSA}.
\end{proposition}

Note that since $\phi$ is a diffeomorphism from $\mathcal{M}^{(1)}$ to $\mathcal{M}^{(2)}$, the probability density function of the manifold $\mathcal{M}^{(2)}$, denoted by $\mu^{(2)}$, can be written as a function of $\mu^{(1)}$ and $\phi$:
\begin{equation}
\mu^{(2)}(y)=\left\vert det\left(\nabla\phi^{-1}(y)\right)\right\vert\mu^{(1)}\left(\phi^{-1}(y)\right)
\end{equation}
where $y\in\mathcal{M}^{(2)}$ and $det()$ denotes the determinant.

The asymptotic expansion of $S_{\epsilon_1,\epsilon_2}$ includes a summation of two Laplace-Beltrami operators (the right term in \eqref{eq:s_exp1}), corresponding to the two considered manifolds, $\mathcal{M}^{(1)}$ and $\mathcal{M}^{(2)}$. 
This term relates to the dynamic Laplacian, defined in \cite{froyland2015dynamic,froyland2017dynamic}, which was shown to be equivalent to the summation of two Laplace-Beltrami operators, from two different time-instances, when assuming a uniform density. The dynamic Laplacian reveals coherent sets in dynamical systems, representing common system behavior in different time-instances. Therefore, this similarity strengthens the claim that the operator $S_{\epsilon_1,\epsilon_2}$ reveals the common structure of the two manifolds.
Conversely, the asymptotic expansion of $A_{\epsilon_1,\epsilon_2}$ is composed of the subtraction between the term \eqref{eq:a_exp1}, which is based on $\mathcal{M}^{(1)}$, and the term \eqref{eq:a_exp2}, which is based on $\mathcal{M}^{(2)}$. These two terms are functions of the probability densities, $\mu^{(\ell)}$ and the diffeomorphism, $\phi$. 
Importantly, in the asymptotic expansion of $A_{\epsilon_1,\epsilon_2}$, the two Laplace-Beltrami operators of the two manifolds that are applied to $f$ in \eqref{eq:s_exp1}, are absent (see Appendix \ref{app:propSA}). Clearly, when $\phi$ is the identity function, i.e. the two manifolds are identical, then $S_{\epsilon_1,\epsilon_2}$ recovers the result in \cite{Coifman2006} and $A_{\epsilon_1,\epsilon_2}$ is zero.

In the following we will show that $A_{\epsilon_1,\epsilon_2}$ characterizes the difference between the manifolds based on differences in their density functions.
In addition, we will show that the eigenfunctions of $A_{\epsilon_1,\epsilon_2}$ are supported on $\Omega^c_{\alpha}$, the regions containing these differences. 
To complement the analysis, in Section \ref{sec:discrete}, we will support these claims in a discrete setting, and in Section \ref{sec:toyexamp} and Section \ref{sec:fECG} we will demonstrate them using both synthetic and real applications.

Consider a special case, where the density $\mu^{(1)}$ of manifold $\mathcal{M}^{(1)}$, is uniform. In this case, the asymptotic expansions in Proposition \ref{prop:SA} reduce to
\begin{align}
S_{\epsilon_1,\epsilon_2} f(x) = & f(x) - \epsilon_1^2\Delta ^{(1)}f(x) - \epsilon_2^2\phi^*\Delta^{(2)}(\phi^*)^{-1}f(x)\label{eq:s_unif_exp1}\\
 & - \frac{\epsilon_2^2}{2}\left( \phi^*\frac{2\nabla^{(2)}(\phi^*)^{-1}f\cdot\nabla^{(2)}\mu^{(2)}}{\mu^{(2)}}(x) - f\phi^*\frac{\Delta^{(2)}\mu^{(2)}}{\mu^{(2)}}(x)\right) + O(\epsilon_1^4+\epsilon_2^4)\\
A_{\epsilon_1,\epsilon_2} f(x) = &  - \frac{\epsilon_2^2}{2}\left( \phi^*\frac{2\nabla^{(2)}(\phi^*)^{-1}f\cdot\nabla^{(2)}\mu^{(2)}}{\mu^{(2)}}(x) + f\phi^*\frac{\Delta^{(2)}\mu^{(2)}}{\mu^{(2)}}(x) \right)  + O(\epsilon_1^4+\epsilon_2^4),
\end{align}
where $\mu^{(2)}(x)=\left\vert det\left(\nabla\phi^{-1}(x)\right)\right\vert$.

In addition, when considering a volume preserving diffeomorphism, similarly to \cite{froyland2015dynamic}, $\mu^{(2)}(x)$ is uniform as well. In such a case, the asymptotic expansion of the operator $S_{\epsilon_1,\epsilon_2}$ is reduced to the addition of the two Laplace-Beltrami operators in \eqref{eq:s_unif_exp1}. Moreover, the second order terms in the asymptotic expansion of $A_{\epsilon_1,\epsilon_2}$ vanish. 
This special case emphasizes that the operator $S_{\epsilon_1,\epsilon_2}$ depends mostly on the geometry of the two manifolds, whereas $A_{\epsilon_1,\epsilon_2}$ depends on the diffeomorphism and the probability density functions of the two manifolds.

\begin{sloppypar}
\begin{proposition}\label{prop:SA_sa}
Denote $\epsilon_1=\epsilon$ and suppose $\epsilon_2=\alpha\epsilon$ for some $\alpha>0$. The operators $A_{\alpha}: C^{\infty}(\mathcal{M}^{(1)}) \rightarrow C^{\infty}(\mathcal{M}^{(1)})$ and $S_{\alpha}: C^{\infty}(\mathcal{M}^{(1)}) \rightarrow C^{\infty}(\mathcal{M}^{(1)})$ are anti-self-adjoint and self-adjoint, respectively,
where $A_{\alpha}=\lim_{\epsilon\rightarrow 0}A_{\epsilon_1,\epsilon_2}/\epsilon^2$ and $S_{\alpha}=\lim_{\epsilon\rightarrow 0}S_{\epsilon_1,\epsilon_2}/\epsilon^2$.
\end{proposition}
The proof is given in Appendix \ref{app:propSA_sa}.
\end{sloppypar}

As presented in this section, the proposed operators, $S_{\epsilon_1,\epsilon_2}$ and $A_{\epsilon_1,\epsilon_2}$, solve the two main shortcomings of the alternating diffusion operator. First, from their asymptotic expansions, it can be seen that there is no dependency on the order of the kernels (this will be revisited in the discrete setting in Section \ref{sec:discrete}). Second, based on Proposition \ref{prop:SA_sa}, they are self-adjoint and anti-self-adjoint, respectively, and therefore, the spectral theorem holds for these operators.

Based on the latter property, we strengthen the claim that $A_{\alpha}$ represents the differences between the two manifolds, by showing that the eigenfunctions of $A_{\alpha}=\lim_{\epsilon\rightarrow 0}A_{\epsilon_1,\epsilon_2}/\epsilon^2$, $\epsilon_2=\alpha\epsilon_1=\alpha\epsilon$, are supported on $\Omega^c_{\alpha}$.
\begin{proposition}\label{prop:Asupp}
Given $f\in C^\infty\left(\mathcal{M}^{(1)}\right)$, if $\mathrm{supp}f\subset\mathring{\Omega}_{\alpha}$, then $A_{\alpha}f(x)=0$.
\end{proposition}
The proof is given in Appendix \ref{app:propAsupp}.
A direct consequence of this proposition is that if $A_{\alpha}f=\lambda f$, $f\neq 0$, then $\mathrm{supp}f\subset\mathcal{M}^{(1)}\backslash\mathring{\Omega}_{\alpha}=\Omega^c_{\alpha}$. Therefore, the eigenfunctions of the difference operator $A_{\alpha}$ (when $\epsilon_1,\epsilon_2\rightarrow 0$) are non-zero only in regions where there are differences between the two manifolds, $\mathcal{M}^{(1)}$ and $\mathcal{M}^{(2)}$.
Note that this proposition does not guarantee the behavior of $f$ on $\Omega^c_{\alpha}$. 

\section{Discrete setting for data analysis\label{sec:discrete}}
We now present our proposed method in the discrete setting. We begin by introducing the discrete counterparts of the operators presented in Section \ref{sec:theory}. In Subsection \ref{sub:disc_interp}, we discuss the differences between the continuous and discrete settings in terms of the diffeomorphism.
In Subsection \ref{sub:disc_sd_prop}, we present our construction of a new coordinate system for the data based on these discrete operators, and in Subsection \ref{sub:disc_op_analysis}, we present a discrete analysis of the operator $A_\epsilon$.

Let $\{x_i\}_{i=1}^N$ and $\{y_i\}_{i=1}^N$ be two datasets of $N$ samples from $\mathcal{M}^{(1)}$ and $\mathcal{M}^{(2)}$, respectively, such that $y_i = \phi(x_i)$. Assume that the data are embedded in two different high dimensional ambient spaces, each corresponding to some measurement of $\mathcal{M}^{(1)}$ or $\mathcal{M}^{(2)}$. Since we only have access to the ambient space, there is no direct access to the geometric structure of $\mathcal{M}^{(1)}$ and $\mathcal{M}^{(2)}$, and identifying $\Omega_{\alpha}$ and $\Omega^c_{\alpha}$ is non-trivial.

Following are the discrete counterparts of the operators presented in Section \ref{sec:theory}.

Let $\mathbf{W}^{(1)}$ and $\mathbf{W}^{(2)}$ be two $N \times N$ affinity (kernel) matrices defined by
\begin{align}
	W_{i,j}^{(1)} = k_{\epsilon_1}^{(1)} (x_i, x_j) \label{eq:W1_disc}\\
	W_{i,j}^{(2)} = k_{\epsilon_2}^{(2)} (y_i, y_j), \label{eq:W2_disc}
\end{align}
where $\epsilon_1,\epsilon_2>0$ may be different, and let $\mathbf{D}^{(1)}$ and $\mathbf{D}^{(2)}$ be two $N \times N$ diagonal matrices, with diagonal elements given by
\begin{align}
	D_{i,i}^{(1)} = \sum \limits _{j=1}^N k_{\epsilon_1}^{(1)} (x_i, x_j)\nonumber\\
	D_{i,i}^{(2)} = \sum \limits _{j=1}^N k_{\epsilon_2}^{(2)} (y_i, y_j).\label{eq:Wmat}
\end{align}

Note that a common choice for $\epsilon_1$ and $\epsilon_2$ in the construction of the kernel used in \eqref{eq:W1_disc}, \eqref{eq:W2_disc} and \eqref{eq:Wmat}, is some scalar multiplication of the median of the distances between the dataset samples, i.e. $\epsilon_1 = c\,\textrm{median}\{d_X(x_i,x_j)\}$ and $\epsilon_2 = c\,\textrm{median}\{d_Y(y_i,y_j)\}$, where $c>0$ is some scalar. By constructing the kernels in this manner, the resulting operators are invariant to scaling between the two underlying manifolds $\mathcal{M}^{(1)}$ and $\mathcal{M}^{(2)}$. Therefore, in the discrete setting, $\Omega_{\alpha}$ is defined similaly to \eqref{eq:defOmega}, by $\Omega_{\alpha}=\left\lbrace x\in\mathcal{M}^{(1)}:\ \nabla\phi\vert_x=\alpha\mathrm{I}\right\rbrace$, where $\alpha>0$ denotes the scaling and $\mathrm{I}$ is the identity matrix.

Let $\mathbf{P}^{(\ell)},\mathbf{Q}^{(\ell)} \in \mathbb{R}^{N \times N}$ be the discrete counterparts of the operators $P^{(\ell)}_{\epsilon_\ell}$ and $Q^{(\ell)}_{\epsilon_\ell}$ given by
\begin{align}
	\mathbf{P}^{(\ell)} &= \left(\mathbf{D}^{(\ell)}\right)^{-1}\mathbf{W}^{(\ell)}\label{eq:Pdisc} \\
	\mathbf{Q}^{(\ell)} &= \mathbf{W}^{(\ell)} \left(\mathbf{D}^{(\ell)}\right)^{-1}\label{eq:Qdisc}
\end{align}
for $\ell=1,2$. It is clear that $\left(\mathbf{P}^{(\ell)}\right)^T = \mathbf{Q}^{(\ell)}$, where $\left(\right)^T$ denotes the transpose operator.
Note that $\mathbf{Q}^{(\ell)}$ is a column stochastic matrix, and therefore, can be interpreted as a Markov transition matrix, defined on the data, which propagates probabilities, analogously to the continuous-time forward operator $Q_{\epsilon_\ell}^{(\ell)}$.

For any $f \in C^{\infty}(\mathcal{M}^{(1)})$, define $\mathbf{v} \in \mathbb{R}^N$ by $v(j) = f(x_j)$. Our formulations are based on the assumption that the discrete matrix and kernel operations approximate the continuous operators, i.e.,
\begin{align}
	P_{\epsilon_\ell}^{(\ell)} f(x_j) &\approx \left(\mathbf{P}^{(\ell)} \mathbf{v}\right)(j) \\
	Q_{\epsilon_\ell}^{(\ell)} f(x_j) &\approx \left(\mathbf{Q}^{(\ell)} \mathbf{v}\right)(j).
\end{align}
This approximation can be justified by a standard large deviation argument, similarly to \cite{singer2016spectral}, which we omit for brevity.

Accordingly, the discrete counterparts of the operators $G_{\epsilon_1,\epsilon_2}$ and $H_{\epsilon_1,\epsilon_2}$ are
\begin{align}
	\mathbf{G} &= \mathbf{P}^{(2)} \mathbf{Q}^{(1)}\label{eq:Gop} \\
	\mathbf{H} &= \mathbf{P}^{(1)} \mathbf{Q}^{(2)}\label{eq:Hop}
\end{align}
and of the operators $S_\epsilon$ and $A_\epsilon$ are
\begin{align}
	\mathbf{S} &= \mathbf{G} + \mathbf{H}\label{eq:Sop} \\
	\mathbf{A} &= \mathbf{G} - \mathbf{H}.\label{eq:Aop}
\end{align}

Note that in this construction of the discrete operators, the probability density function of each manifold, $\mu^{(\ell)}$, is reflected in the sampling of the points in the dataset.
In addition, we assume that the diffeomorphism $\phi$, which appears explicitly in the continuous operators, $S_{\epsilon_1,\epsilon_2}$ and $A_{\epsilon_1,\epsilon_2}$, is implicitly contained in the discrete operators $\mathbf{P}$ and $\mathbf{Q}$.

\begin{proposition}\label{prop:SA_sym_discrete}
$\mathbf{S}$ is symmetric and $\mathbf{A}$ is anti-symmetric.
\end{proposition}
Based on the definitions of $\mathbf{S}$ and $\mathbf{A}$ above, it is easy to show that $\mathbf{S}^T=\mathbf{S}$ and that $\mathbf{A}^T=-\mathbf{A}$. Specifically, $(j\mathbf{A})^H=j\mathbf{A}$, where $\left(\right)^H$ denotes conjugate transpose and $j=\sqrt{-1}$.

Note that both the discrete alternating diffusion operator \cite{lederman2015learning} and the operators $\mathbf{G}$ and $\mathbf{H}$, are not Hermitian and therefore, there is no spectral decomposition for them. Moreover, by their definition, they depend on the order of the matrix multiplication, e.g. whether we define $\mathbf{G}=\mathbf{P}^{(2)}\mathbf{Q}^{(1)}$ or $\mathbf{G}=\mathbf{P}^{(1)}\mathbf{Q}^{(2)}$.

The use of the symmetric and anti-symmetric parts of an operator in the context of constructing a new representation was also presented in \cite{fanuel2018magnetic}, where representations for directed graphs were obtained based on the symmetric and anti-symmetric parts of the non-symmetric weight matrix of the graph.

\subsection{Interpretation of the operators and diffeomorphism in the discrete setting\label{sub:disc_interp}}
Note that in the current definition of the discrete operators $\mathbf{S}$ and $\mathbf{A}$, we apply operators defined on $\mathcal{M}^{(1)}$ and operators defined on $\mathcal{M}^{(2)}$ to the same functions. 
Specifically, applying $\mathbf{H}$ to $\mathbf{v}$, a discretization of $f\in C^\infty\left(\mathcal{M}^{(1)}\right)$, implies that the function $f$ is first pushed forward to $\mathcal{M}^{(2)}$ and then discretized. Namely, the discrete operators, $\mathbf{G}$ and $\mathbf{H}$, embody both the continuous operators, $G_\epsilon$ and $H_\epsilon$, respectively, and the diffeomorphism, $\phi$.
When the two datasets significantly differ in their densities or metrics, this could be incorrect. One option to solve this is by defining the following operators
\begin{align}
	\tilde{\mathbf{S}} &= \mathbf{Q}^{(1)} \mathbf{S} \mathbf{P}^{(1)} \label{eq:Stilde} \\
	\tilde{\mathbf{A}} &= \mathbf{Q}^{(1)} \mathbf{A} \mathbf{P}^{(1)}. \label{eq:Atilde}
\end{align}
These operators are symmetric and anti-symmetric, respectively, and preserve the same asymptotic behavior. 
A second option is to use concepts from \cite{ovsjanikov2012functional}, which presents a method for recovering a functional map between two shapes, and include such a functional map, between the two manifolds, in the construction of the operators $\mathbf{S}$ and $\mathbf{A}$. 
We note that in the experimental results, presented in Section \ref{sec:toyexamp} and Section \ref{sec:fECG}, both operator forms $\tilde{\mathbf{S}}$, $\tilde{\mathbf{A}}$, and $\mathbf{S}$, $\mathbf{A}$, led to comparable results. This is due to the similarity of the two manifolds in these applications.

\subsection{New representations of the data based on $\mathbf{S}$ and $\mathbf{A}$\label{sub:disc_sd_prop}}
Our goal is to obtain new representations for multi-modal data based on the operators $\mathbf{S}$ and $\mathbf{A}$, analogous to the diffusion maps coordinates \cite{Coifman2006} that represent the diffusion distances in the data. 
Specifically, we seek non-linear mappings of the data to new coordinate systems, which describe the common structures or the differences between the modalities (manifolds). 
In addition, to obtain a compact representation, we require the constructed coordinates to be orthogonal. In this subsection, we present one option for obtaining such representations.

Since $\mathbf{S}$ is a symmetric matrix, it has real eigenvalues and eigenvectors. The eigenvectors are orthogonal, and hence, we can construct a new low-dimensional representation for the common structures in the datasets based on $\mathbf{S}$, by taking its eigenvectors, corresponding to the largest eigenvalues.

The operator $\mathbf{A}$ is anti-symmetric, and therefore, has purely imaginary eigenvalues, in conjugate pairs, and complex eigenvectors. In order to construct a new low-dimensional representation for the differences between the datasets based on $\mathbf{A}$, we show in the following that by taking the real and imaginary parts of non-conjugate eigenvalues, we obtain a set of orthogonal vectors. Therefore, we propose to construct a new representation based on $\mathbf{A}$, by taking the real and imaginary parts (separately) corresponding to the largest (in absolute value) non-conjugate eigenvalues.

The spectral decomposition of a real anti-symmetric matrix is given by
\begin{equation}
\mathbf{A} = \Psi\Lambda\Psi^T,\label{eq:A_disc_decomp}
\end{equation}
where $\Psi$ is a matrix containing the eigenvectors of $\mathbf{A}$ in its columns and $\Lambda$ is a diagonal matrix, containing the eigenvalues in conjugate pairs, i.e.:
\begin{equation}
\begin{bmatrix}
j\lambda_1 & 0  & 0 & \ldots\\
0 & -j\lambda_1 & 0 & \ldots\\
0 & 0 & j\lambda_2 & \ldots\\
0 & 0 & 0 & \ddots
\end{bmatrix},
\end{equation}
where $\lambda_k$, $k=1,\ldots,\lceil N/2\rceil$,  are real and positive and $j$ denotes $\sqrt{-1}$. Note that when $N$ is odd, $\lambda_{\lceil N/2\rceil}=0$.

This spectral decomposition is related to a real orthogonal decomposition of the form:
\begin{equation}
\mathbf{A} = U\Sigma U^T,
\end{equation}
where $U$ is orthogonal and real, and $\Sigma$ is a block diagonal matrix, with $k$-th $2\times 2$ diagonal blocks of the form:
\begin{equation}
\begin{bmatrix}
0 & \mu_k\\
-\mu_k & 0
\end{bmatrix},
\end{equation}
where $\mu_k=\lambda_k$ for $k=1,\ldots, \lfloor N/2\rfloor$ \cite{gantmakher1998theory}. 
By comparing this form to the spectral decomposition of the anti-symmetric matrix, $\mathbf{A}$, it can be shown that the real and imaginary parts of eigenvectors corresponding to non-conjugate nonzero eigenvalues of $\mathbf{A}$, are equal to different orthogonal vectors in $U$, i.e. $\mathrm{real}\{\psi_{\ell}\}=u_k$, $\mathrm{imag}\{\psi_{\ell}\}=u_n$, where $\psi_{\ell}$ is the $\ell$'th eigenvector of $\mathbf{A}$ and $u_k$ and $u_n$ are the $k$-th and $n$-th columns of $U$ ($n\neq k$). From the orthogonality of $U$, we obtain $\left\langle \mathrm{real}\{\psi_{\ell}\},\mathrm{imag}\{\psi_{r}\} \right\rangle=0$ \ $\forall \lambda_{\ell}\neq \lambda_r$ or for the real and imaginary parts of the same eigenvector ($\ell=r$). 
The real and imaginary parts of these eigenvectors can then be used for the construction of a new orthogonal representation for the differences between the datasets.

Algorithm \ref{alg:SA} summarizes the procedure for obtaining the new representations for the data based on $\mathbf{S}$ and $\mathbf{A}$.

\begin{algorithm}
\caption{Representation of the common structures and the differences between datasets\label{alg:SA}}
\begin{enumerate}
\item Construct the affinity matrices for the two datasets:
\begin{align}
W_{i,j}^{(1)}=\exp\left(-\frac{d_X(x_i,x_j)^2}{\epsilon_1^2}\right), & & W_{i,j}^{(2)}=\exp\left(-\frac{d_Y(y_i,y_j)^2}{\epsilon_2^2}\right),\label{eq:kernConst}
\end{align}
where $d_X, d_Y$ are some notion of distance, defined on the data (e.g. the Euclidean distance if the data are in an ambient Euclidean space) and $\epsilon_1^2, \epsilon_2^2$ are the kernel scales, commonly taken as some multiplication of the median of the distances.
\item Create the row stochastic and column stochastic matrices:
\begin{align}
\mathbf{P}^{(1)}=\left(\mathbf{D}^{(1)}\right)^{-1}\mathbf{W}^{(1)}, & & \mathbf{Q}^{(1)}=\mathbf{W}^{(1)}\left(\mathbf{D}^{(1)}\right)^{-1}\nonumber\\
\mathbf{P}^{(2)}=\left(\mathbf{D}^{(2)}\right)^{-1}\mathbf{W}^{(2)}, & & \mathbf{Q}^{(2)}=\mathbf{W}^{(2)}\left(\mathbf{D}^{(2)}\right)^{-1},\label{eq:alg1_PQ}
\end{align}
where $\mathbf{D}^{(\ell)}$ is a diagonal matrix with $D_{i,i}^{(\ell)}=\sum_{j=1}^{N}W_{i,j}^{(\ell)}$ and $\ell=1,2$.
\item Construct the symmetric and anti-symmetric matrices:
\begin{align}
\mathbf{S}=\mathbf{P}^{(2)}\mathbf{Q}^{(1)} + \mathbf{P}^{(1)}\mathbf{Q}^{(2)}, & & \mathbf{A}=\mathbf{P}^{(2)}\mathbf{Q}^{(1)} - \mathbf{P}^{(1)}\mathbf{Q}^{(2)}.\label{eq:alg1_SA}
\end{align}
\item To obtain a new representation of dimension $M$ for the common structures in the two datasets, calculate the eigenvalue decomposition of $\mathbf{S}$, $\mathbf{S}\psi_k^\mathbf{S}=\lambda_k^\mathbf{S}\psi_k^\mathbf{S}$, and take the first $M$ eigenvectors, corresponding to the largest eigenvalues, $\{x_i,y_i\}\mapsto\{\psi^\mathbf{S}_k(i)\}_{k=1}^M$.
\item To obtain a representation for the differences between the datasets, calculate the eigenvalue decomposition of $\mathbf{A}$, $\mathbf{A}\psi_k^\mathbf{A}=\lambda_k^\mathbf{A}\psi_k^\mathbf{A}$, and take the real and imaginary parts of the first $M/2$ eigenvectors, corresponding to the largest (in absolute value) non-conjugate eigenvalues, $\{x_i,y_i\}\mapsto\{\mathrm{real}\{\psi^\mathbf{A}_k(i)\};\mathrm{imag}\{\psi^\mathbf{A}_k(i)\}\}_{k=1}^{M/2}$.
\end{enumerate}
\end{algorithm}

\subsection{Discrete analysis of the operator $\mathbf{A}$\label{sub:disc_op_analysis}}
In this subsection we present an analysis for the discrete operator $\mathbf{A}$, showing that it is supported on the locations of the differences between the datasets, similarly to the continuous operator $A_\epsilon$.

Consider two datasets $\{x_i\}_{i=1}^N$ and $\{y_i\}_{i=1}^N$, each consisting of $N$ points, which are samples of $\mathcal{M}^{(1)}$ and $\mathcal{M}^{(2)}$ respectively. 
The affinity matrices for the datasets $\{x_i\}$ and $\{y_i\}$ are constructed according to \eqref{eq:kernConst} and are denoted by $\mathbf{W}^{(1)}$ and $\mathbf{W}^{(2)}$, respectively. 
Define $V_{\Omega_{\alpha}}=\{i,j\in V \ | \ W^{(1)}_{i,j}=W^{(2)}_{i,j}\}$, where $V=\{1,\dots,N\}$, and $V_{\Omega^c_{\alpha}}=V\backslash V_{\Omega_{\alpha}}$.
Assume that the correspondence between pairs of points in $\{x_i\}$ and $\{y_i\}$ is given and that the datasets differ in the affinities between the points, i.e. $W_{i,j}^{(1)}\neq W_{i,j}^{(2)}$ if and only if $i\in V_{\Omega^c_{\alpha}}$ and $j\in V_{\Omega^c_{\alpha}}$. 
Note that this can be approximated by choosing small kernel scales, $\epsilon_x^2,\epsilon_y^2$, in \eqref{eq:kernConst}, such that the effect of the differences is localized around them. This indicates that the choice of $\epsilon$ is important in the construction of $\mathbf{A}$ and should be smaller than the median of the distances in the data, which is common practice.

\begin{proposition}
Suppose $|V_{\Omega^c_{\alpha}}|=m\leq N/2$. The discrete operator $\mathbf{A}$ has the following properties:
\begin{enumerate}
\item $A_{i,j}\neq 0$ only when $i\in V_{\Omega^c_{\alpha}}$ or $j\in V_{\Omega^c_{\alpha}}$.
\item The rank of $\mathbf{A}$ is bounded by $2\left|V_{\Omega^c_{\alpha}}\right|=2m$
\end{enumerate}
\end{proposition}

This proposition states that the discrete operator $\mathbf{A}$ is non-zero only in regions where the two datasets differ and that its rank is related to the dimensionality of the differences. A direct consequence of this proposition is that the eigenvectors of $\mathbf{A}$ encode information related to the location of the non-trivial diffeomorphism, and hence, $\mathbf{A}$ can be utilized for representing the differences between the two datasets.

\begin{proof}
Based on the definition of the datasets and the assumptions stated above, the difference between the affinity matrices $\mathbf{W}^{(1)}, \mathbf{W}^{(2)}\in\mathbb{R}^{N\times N}$ can be represented by
\begin{equation}
\mathbf{W}^{(2)}=\mathbf{W}^{(1)}+\mathbf{B}^T \mathbf{B},
\end{equation}
where $\mathbf{B}\in\mathbb{R}^{N\times N}$, $\mathbf{B}e_i =0$ if $i\in V_{\Omega_{\alpha}}$ and $e_i$ are vectors which contain $1$ at index $i$ and $0$ elsewhere, i.e., $\mathbf{B}$ is a matrix in which column $i$ contains only $0$ $\forall i\in V_{\Omega_{\alpha}}$. Note that $(\mathbf{B}^T\mathbf{B})_{i,j}=0$ if $i\in V_{\Omega_{\alpha}}$ or if $j\in V_{\Omega_{\alpha}}$.

Then, based on the definition of $\mathbf{D}^{(\ell)}$ in \eqref{eq:Wmat}, $\mathbf{D}^{(2)}=\mathbf{D}^{(1)} + \mathrm{diag}\left(\mathbf{B}^T\mathbf{B}\mathbf{1}\right)$, where $\mathbf{1}\in\mathbb{R}^N$ is a vector containing only $1$'s and $\mathrm{diag}(z)$ denotes a diagonal matrix with the elements of $z$ on its diagonal. Note that $\mathrm{diag}\left(\mathbf{B}^T\mathbf{B}\mathbf{1}\right)$ is a diagonal matrix with non-zero diagonal entries only for $i\in V_{\Omega^c_{\alpha}}$.

The operator $\mathbf{P}^{(2)}$ is then given by
\begin{equation}
\mathbf{P}^{(2)}=\left(\mathbf{D}^{(2)}\right)^{-1}\mathbf{W}^{(2)}=\left(\mathbf{D}^{(1)}+\mathrm{diag}\left(\mathbf{B}^T\mathbf{B}\mathbf{1}\right)\right)^{-1}\left(\mathbf{W}^{(1)}+\mathbf{B}^T\mathbf{B}\right).
\end{equation}
Denote the inverse of the sum $\left(\mathbf{D}^{(1)}+\mathrm{diag}\left(\mathbf{B}^T\mathbf{B}\mathbf{1}\right)\right)$ by $\left(\left(\mathbf{D}^{(1)}\right)^{-1} - \mathbf{J}\right)$, where
\begin{equation}
\mathbf{J}=\left(\mathbf{I}+\left(\mathbf{D}^{(1)}\right)^{-1}\mathrm{diag}\left(\mathbf{B}^T\mathbf{B}\mathbf{1}\right)\right)^{-1}\left(\mathbf{D}^{(1)}\right)^{-1}\mathrm{diag}\left(\mathbf{B}^T\mathbf{B}\mathbf{1}\right)\left(\mathbf{D}^{(1)}\right)^{-1}\label{eq:Jmat}
\end{equation} 
and $\mathbf{I}$ denotes the identity matrix.
Based on the expression in \eqref{eq:Jmat}, $\mathbf{J}$ is a diagonal matrix with non-zero values only for $i\in V_{\Omega^c_{\alpha}}$, i.e., $J_{i,i}=0 \ \ \forall i\not\in V_{\Omega^c_{\alpha}}$.

Substituting these derivations into the definition of $\mathbf{A}$ we get
\begin{eqnarray}
\mathbf{A} & = & \mathbf{P}^{(1)}\left(\mathbf{P}^{(2)}\right)^T-\mathbf{P}^{(2)}\left(\mathbf{P}^{(1)}\right)^T\\
& = & \mathbf{P}^{(1)}\left(\mathbf{B}^T\mathbf{B}\left(\mathbf{D}^{(1)}\right)^{-1} - \mathbf{W}^{(1)}\mathbf{J}-\mathbf{B}^T\mathbf{B}\mathbf{J}\right)\\
& & -\left(\left(\mathbf{D}^{(1)}\right)^{-1}\mathbf{B}^T\mathbf{B} - \mathbf{J}\mathbf{W}^{(1)} -\mathbf{J}\mathbf{B}^T\mathbf{B}\right)\left(\mathbf{P}^{(1)}\right)^T.
\end{eqnarray}
Since all the elements in this expression are multiplied either by $\mathbf{J}$ or $\mathbf{B}^T\mathbf{B}$, which contain non-zero values only for rows and columns corresponding to $i\in V_{\Omega^c_{\alpha}}$, the value of the discrete operator are $A_{i,j}\neq 0$ only when $i\in V_{\Omega^c_{\alpha}}$ or $j\in V_{\Omega^c_{\alpha}}$. In addition, this indicates that the rank of $\mathbf{A}$ is bounded by $2|V_{\Omega^c_{\alpha}}|=2m$, i.e., twice the number of elements in $V_{\Omega^c_{\alpha}}$.
\end{proof}

\section{Non-isometric shapes analysis\label{sec:toyexamp}}

In this section, we demonstrate the properties of the proposed operators $\mathbf{S}$ and $\mathbf{A}$ using a toy example composed of two manifolds with a non-isometric diffeomorphism. We show that operator $\mathbf{S}$ recovers the common manifold, whereas operator $\mathbf{A}$ captures the ``difference'' between the two manifolds. 

\begin{sloppypar}
Consider two manifolds, $\mathcal{M}^{(1)}$, which is a sphere, and $\mathcal{M}^{(2)}$, which is a sphere with scaling and a non-isometric deformation, which we will refer to as a ``bump''. The two manifolds (shapes) are depicted in Figure \ref{fig:Spheres}. In this example, $\Omega_{\alpha}$, defined in \eqref{eq:defOmega}, represents the part of the sphere that does not undergo deformation, and $\Omega^c_{\alpha}$, defined in \eqref{eq:defOmegaC}, represents the deformed part.
In order to construct the operators $\mathbf{S}$ and $\mathbf{A}$, we first construct the two diffusion operators, $\mathbf{P}^{(\ell)}$ and $\mathbf{Q}^{(\ell)}$, for each manifold $\ell=1,2$, as described in \eqref{eq:alg1_PQ}. 
We then construct the symmetric and anti-symmetric discrete operators, $\mathbf{S}$ and $\mathbf{A}$, respectively, according to \eqref{eq:Sop} and \eqref{eq:Aop}. 
Finally, the eigenvalue decompositions of $\mathbf{S}$ and $\mathbf{A}$ are calculated, and the eigenvectors are sorted according to the imaginary part of the eigenvalues in descending order.
The kernel scales, $\epsilon_1$ and $\epsilon_2$ in \eqref{eq:kernConst}, were set to be the median of the distances, $d_X(x_i,x_j)$ and $d_Y(y_i,y_j)$ respectively, divided by some scalar. In the construction of $\mathbf{S}$, $\epsilon_1$ was set to $median(d_X(x_i,x_j))/2$ and $\epsilon_2$ to $median(d_Y(y_i,y_j))/2$, and in the construction of $\mathbf{A}$, $\epsilon_1$ was set to $median(d_X(x_i,x_j))/5$ and $\epsilon_2$ to $median(d_Y(y_i,y_j))/5$. The choice of smaller kernel scales in the construction of the operator $\mathbf{A}$ was motivated by the discrete analysis presented in Subsection \ref{sub:disc_op_analysis}.
\end{sloppypar}

\begin{figure}
\centering
\subfloat[]{\includegraphics[width=0.32\textwidth]{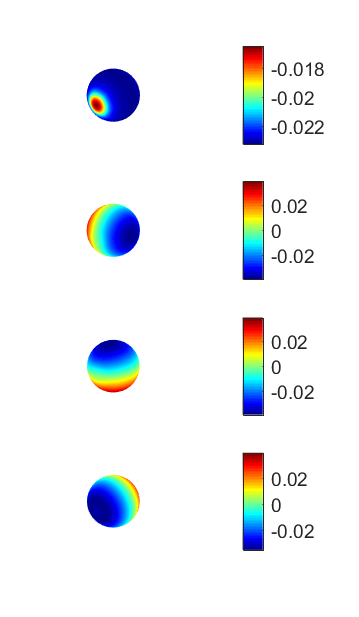}
}\subfloat[]{\includegraphics[width=0.32\textwidth]{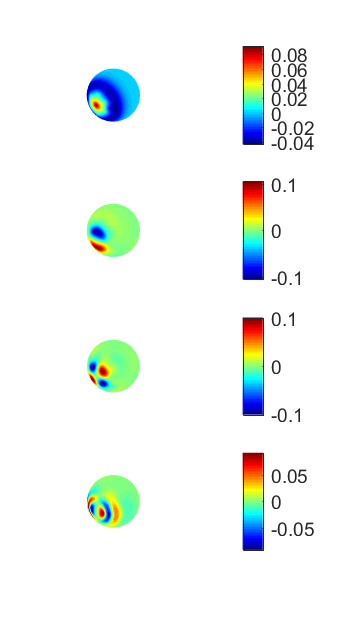}
}\subfloat[]{\includegraphics[width=0.32\textwidth]{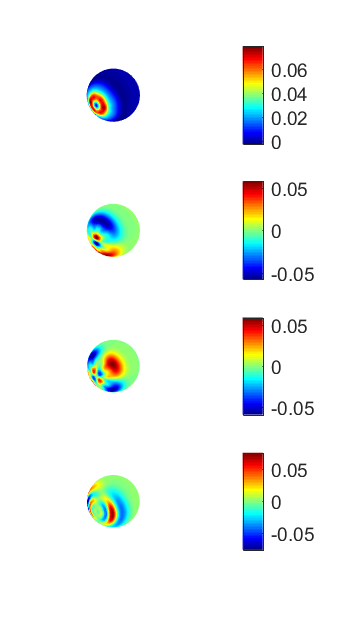}
}

\subfloat[]{\includegraphics[width=0.32\textwidth]{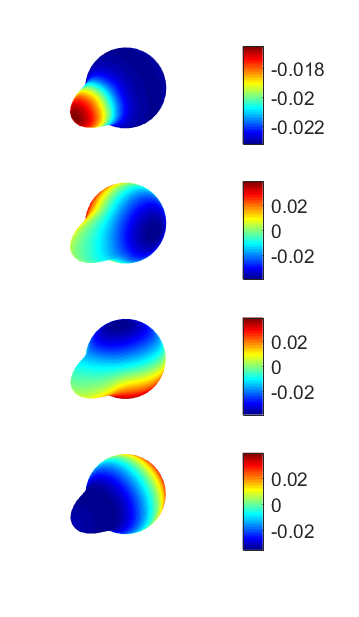}
}\subfloat[]{\includegraphics[width=0.32\textwidth]{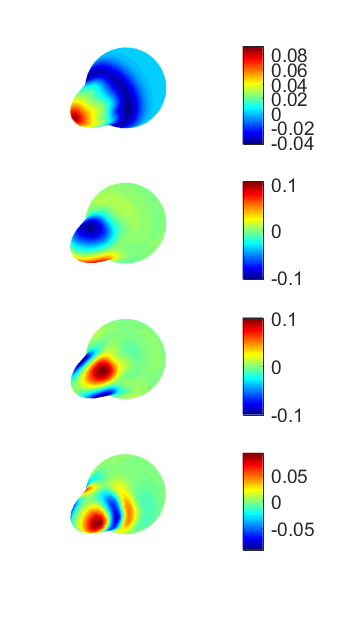}
}\subfloat[]{\includegraphics[width=0.32\textwidth]{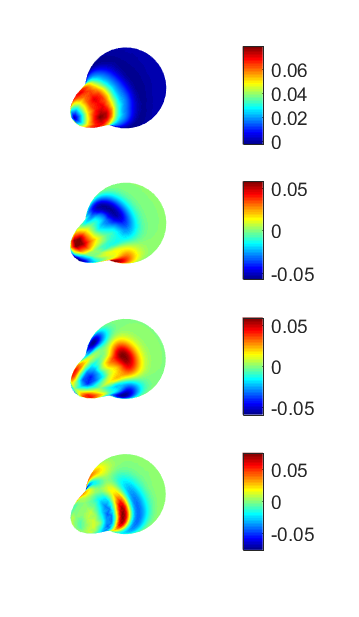}
}

\caption{Application of operators $\mathbf{S}$ and $\mathbf{A}$ to two diffeomorphic manifolds, a sphere (plots (a)-(c)) and a deformed sphere (plots (d)-(f)). The plots in the first row depict the sphere colored by the: (a) the eigenvectors of the symmetric operator $\mathbf{S}$, (b) the real part of the eigenvectors of the anti-symmetric operator $\mathbf{A}$, (c) the imaginary part of the eigenvectors of the anti-symmetric operator. The coloring of plots (d)-(f) corresponds to the coloring of plots (a)-(c).\label{fig:toyEx}}
\end{figure}

Figure \ref{fig:toyEx} presents the sphere (top plots) and the bump (bottom plots), colored by the eigenvectors of the operators $\mathbf{S}$ and $\mathbf{A}$. 
Plots (a) and (d) are colored by the first $4$ eigenvectors of $\mathbf{S}$. Plots (b) and (e) are colored by the real part of the first $4$ eigenvectors of $\mathbf{A}$, and plots (c) and (f) are colored by the imaginary part of the first $4$ eigenvectors of $\mathbf{A}$. 
Note that in both $\mathbf{S}$ and $\mathbf{A}$, the eigenvector corresponding to the largest eigenvalue (top plot) separates between the location of the deformation and the similar parts of the sphere.
The other $3$ eigenvectors of $\mathbf{S}$ and $\mathbf{A}$ exhibit different properties. The eigenvectors of $\mathbf{S}$ are supported on the entire sphere and take the form of standard spherical harmonics. 
Conversely, the eigenvectors of $\mathbf{A}$ (both real and imaginary parts) are supported on the deformed part $\Omega^c_{\alpha}$ and take the form of local standard spherical harmonics there. 
Namely, the eigenvectors of $\mathbf{A}$ are supported on the regions where the diffeomorphism is non-isometric, and within their support, the ``standard'' (yet, local) harmonic oscillations are obtained. 
 
\section{Fetal ECG\label{sec:fECG}}
In this section, we demonstrate the properties of the proposed operators in a fetal heart activity identification problem, from two trans-abdominal maternal ECG (ta-mECG) contacts. This problem consists of two oscillatory signals, one is the undesired maternal ECG signal and the other is the desired fetal ECG signal. The two signals are observed by two ECG contacts located on the maternal abdomen. In each contact, a mixture of the two oscillatory signals is captured. Based on the physiological properties, we assume that both observations capture the same view of the maternal ECG signal, since the source of the maternal ECG is located remotely from the two abdominal contacts. Conversely, we assume that the two observations capture different views of the fetal ECG signal, since its source is located close to each of the contacts.

Fetal heart rate (fHR) provides significant information about fetal health. For example, fetal distress monitoring can be obtained through fHR analysis \cite{jenkins1989thirty}. 
In recent years, analyzing how fHR fluctuates has attracted increasing attention due to its potential to enhance our understanding of the dynamics of various physiological systems, as well as to contribute to clinical procedures, e.g., inflammation detection \cite{durosier2015does}. 
Obtaining intrapartum fHR non-invasively is not an easy task. Traditionally, cardiotocogram is the standard tool to obtain the fHR. However, it has been well known that the fHR obtained by cardiotocogram does not have a sufficiently high sampling rate for the fHR fluctuation analysis. 
In the past decades, studies have focused on obtaining the fHR through the ta-mECG, due to the high sampling rate of the ECG. See, for example \cite{andreotti2016open,li2017efficient}, and references therein. 
However, to date, while many algorithms and products based on multiple channels (more than 4) have been proposed, there is no gold-standard that works in all situations when there are only one or two channels.
While we do not presume to provide a state-of-the-art algorithm, in this section, we show the potential of the operator $\mathbf{A}$ in extracting the fHR from two ta-mECG signals.

The section is structured as follows. In subsection \ref{sub:fECGmodel}, we present our basic geometric model of the problem, to justify the application of the operator $\mathbf{A}$.
Results on simulation data are presented in subsection \ref{sub:semirealapp} and on real measured data in subsection \ref{sub:realapp}.

\subsection{Model\label{sub:fECGmodel}}
Let $s^{(\ell)}(t)$, $\ell=1,2$, be the measured signal at the first and second ta-mECG leads, given by
\begin{align*}
	s^{(\ell)}(t) = m^{(\ell)}(t)+f^{(\ell)}(t),
\end{align*}
where $f^{(\ell)}(t)$ and $m^{(\ell)}(t)$ denote the fetal and maternal ECG signals, respectively. The signal $f^{(\ell)}(t)$ (resp. $m^{(\ell)}(t)$) consists of a (quasi) periodic oscillation representing the fetal (resp. the maternal) heart beat. 
``Quasi'' here indicates that the heart rate and ECG morphology change occasionally.
To simplify the discussion, we assume that the relationship between the two (separate) cardio systems entails that the maternal and fetal ECG signals are approximately perpendicular in short time periods, i.e.,
\begin{equation}\label{eq:ecg_assumption}
	\int _{I} m^{(\ell)}(t) f^{(\ell)}(t) dt \approx 0
\end{equation}
for all time intervals $I$ of length $1$ second.
Note that this is an over-simplified model motivated by the fact that the maternal hear rate is about $1Hz$ and the fetal and maternal heart beats are not synchronized. Indeed, when the QRS complexes of the maternal and fetal ECG overlap, this assumption may not hold.

Using lag map embedding, the measured signals can be written as
\begin{align*}
	\mathbf{s}^{(\ell)}(t) & = \mathbf{m}^{(\ell)}(t) + \mathbf{f}^{(\ell)}(t) \in \mathbb{R}^p,
\end{align*}
where $\mathbf{s}^{(\ell)}(t)=\left[s^{(\ell)}(t),Ts^{(\ell)}(t),...,T^{p-1} s^{(\ell)}(t)\right]$, $T$ denotes an operator that propagates $s^{(\ell)}(t)$ one time step forward and $p$ is the number of time steps in the lag map embedding of each time interval $I$.

Let $\mathcal{E}^{(1)} \subset \mathbb{R}^p$ and $\mathcal{E}^{(2)} \subset \mathbb{R}^p$ be the embedding of $\mathbf{s}^{(1)}(t)$ and $\mathbf{s}^{(2)}(t)$ in $I$, respectively. By assumption \eqref{eq:ecg_assumption}, we can write
\begin{align*}
	\mathcal{E}^{(1)} = \mathcal{M}^{(1)} \oplus \mathcal{F}^{(1)},
\end{align*}
where $\mathcal{F}^{(1)}$ and $\mathcal{M}^{(1)}$ are the manifolds underlying $\mathbf{f}^{(1)}(t)$ and $\mathbf{m}^{(1)}(t)$ in $I$, respectively.
Similarly, let $\mathcal{F}^{(2)}$ and $\mathcal{M}^{(2)}$ be the manifolds underlying $\mathbf{f}^{(2)}(t)$ and $\mathbf{m}^{(2)}(t)$ in $I$, respectively.

As described at the beginning of Section \ref{sec:fECG}, the locations of the two abdominal leads entail that $\mathbf{m}^{(1)}(t) \approx \mathbf{m}^{(2)}(t)$ while $\mathbf{f}^{(1)}(t)$ and $\mathbf{f}^{(2)}(t)$ are different. As a result, $\mathcal{M}^{(1)} \approx \mathcal{M}^{(2)}$ and the diffeomorphism between $\mathcal{E}^{(1)}$ and $\mathcal{E}^{(2)}$ can be modeled as
\begin{equation*}
	\mathcal{E}^{(2)} = \phi(\mathcal{E}^{(1)}) = \phi(\mathcal{M}^{(1)} \oplus \mathcal{F}^{(1)}) = \mathcal{M}^{(1)} \oplus \tilde{\phi}(\mathcal{F}^{(1)}),
\end{equation*}
where $\tilde{\phi}:\mathcal{F}^{(1)} \rightarrow \mathcal{F}^{(2)}$ is a smooth diffeomorphism.

Define $\mu^{(\ell)}$ as the probability density on $\mathcal{E}^{(\ell)}$, $\mu^{(\ell)}_m$ as the marginal density of $\mu^{(\ell)}$ on $\mathcal{M}^{(\ell)}$ and $\mu^{(\ell)}_f$ as the marginal density of $\mu^{(\ell)}$ on $\mathcal{F}^{(\ell)}$.

\begin{sloppypar}
\begin{corollary}\label{cor:Asupp}
Define $A_{\alpha} = \lim_{\epsilon\rightarrow 0} A_{\epsilon_1,\epsilon_2}/\epsilon^2$, where $\epsilon_2=\alpha\epsilon$ and $\epsilon_1=\epsilon$, $\alpha>0$.
For all $g\in C^{\infty}(\mathcal{E}^{(1)})$, if $\textmd{supp}g \subset \mathcal{M}^{(1)} \oplus \mathring{\Omega}_{f,\alpha}$, then $A_{\alpha}g = 0$.
Hence, if $A_{\alpha}g = \lambda g$, $g \neq 0$, then, $\textmd{supp}g \subset \mathcal{M}^{(1)} \oplus \Omega_{f,\alpha}^c$.
\end{corollary}
\end{sloppypar}

According to this corollary, the eigenfunctions of the operator $A_{\alpha}$ are supported on the differences. We assume that the differences in the measured fetal ECG signals are manifested mainly during heart activity, i.e. depolarization (QRS complex and P wave) and re-polarization (T wave). Therefore, based on the model presented in this subsection, the eigenfunctions of $A_{\alpha}$ can serve as indicators for fetal heart activity.
In addition, the common component in this model, i.e. the maternal heart activity, $\mathbf{m}^{(1)}(t)$ and $\mathbf{m}^{(2)}(t)$, represents an almost periodic oscillation. By the Takens' embedding theorem, the manifolds underlying such signals can be well recovered, up to a diffeomorphism, by a 1-dimensional manifold that is diffeomorphic to $\mathcal{S}^1$. Therefore, we expect that the eigenfunctions of the operator $S_{\alpha}$ will represent $\mathcal{S}^1$.

\subsection{Fetal heart rate detection -- synthetic example \label{sub:semirealapp}}
In this subsection, we begin with a synthetic problem setting of fetal ECG detection to demonstrate the main properties of our composite operators for such data. 

\begin{figure}
\centering
\includegraphics[width=0.9\textwidth]{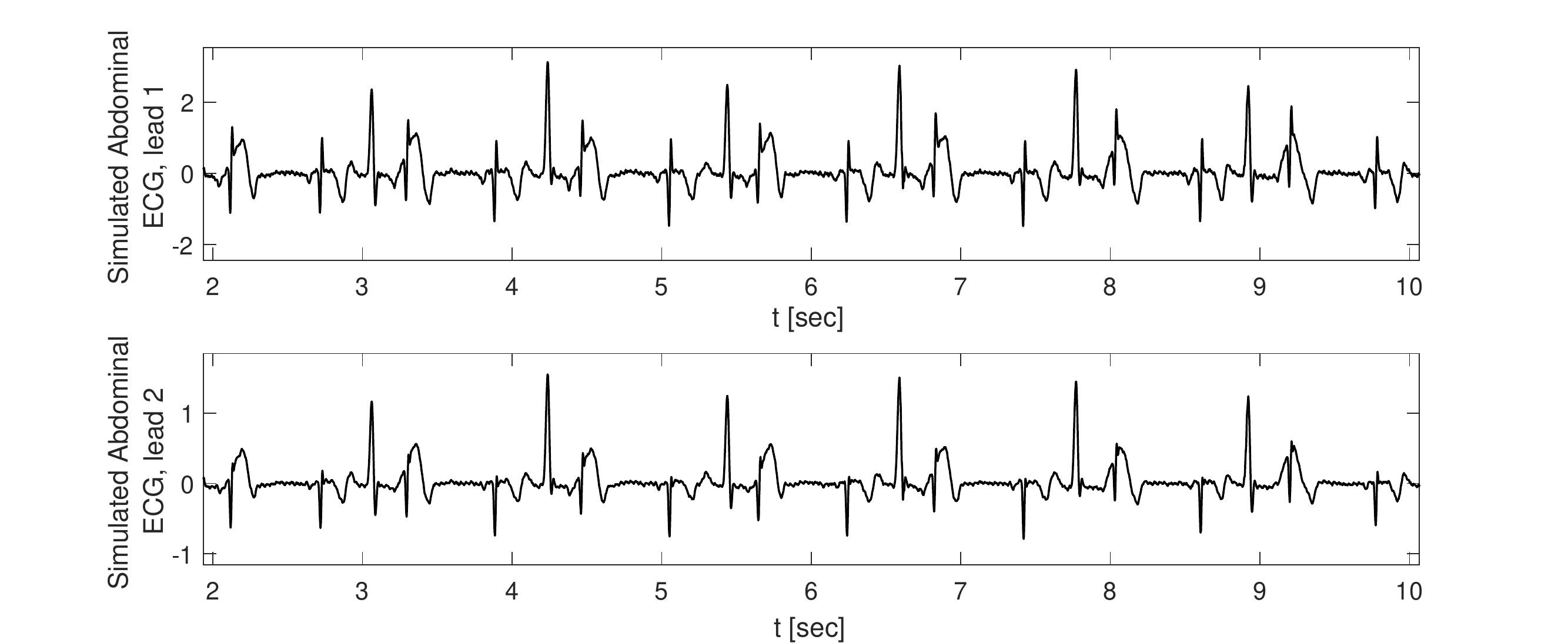}
\caption{Two simulated ECG leads, representing two ta-mECG recordings\label{fig:syntData}}
\end{figure}

Following the model described in Subsection \ref{sub:fECGmodel}, we create synthetic data of two ta-mECG leads from three different ECG recordings, denoted by $z^{(1)}_i$, $z^{(2)}_i$ and $z^{(3)}_i$, where $i=1,\dots,N$ and $N=4\times 10^4$ is the number of samples. 
These recordings are taken from the QT database in Physionet \cite{laguna1997database,PhysioNet}, which contains annotated 2-lead ECG recordings, sampled at $250Hz$. The signals $z^{(2)}_i$ and $z^{(3)}_i$ are taken from the same recording, i.e. taken from two corresponding ECG leads which were recorded simultaneously. These recordings were filtered by a notch filter to remove the $60Hz$ net noise and by a median filter (with a window size of $100$ samples) to remove the baseline drift. 
In order to obtain more samples per heart cycle, we increased the number of samples in these recordings using interpolation. 
One recording, $z^{(1)}_i$, represents the maternal ECG, and is upsampled by a factor of $4$.
The other two recordings, $z^{(2)}_i$ and $z^{(3)}_i$, represent the fetal ECG, which commonly has a higher heart rate, and therefore, they are upsampled by a factor of $2$.
The simulated ta-mECG signals $s^{(1)}_i$ and $s^{(2)}_i$ are generated according to
\begin{eqnarray}
s^{(1)}_i & = & 2z^{(1)}_i - z^{(2)}_i\\
s^{(2)}_i & = & z^{(1)}_i  - 0.5z^{(3)}_i.
\end{eqnarray}
where the common maternal ECG $z^{(1)}_i$ is identical up to a scaling factor.
In these simulated signals, $z^{(1)}_i$, which is denoted by $m^{(\ell)}(t)$ in Subsection \ref{sub:fECGmodel}, is assumed to be part of the common structure, whereas the fetal ECG signals $z^{(2)}_i$ and $z^{(3)}_i$ are captured differently by the two abdominal leads. With regard to the model described in Subsection \ref{sub:fECGmodel}, the fetal ECG signals $z^{(2)}_i$ and $z^{(3)}_i$, denoted there by $f^{(\ell)}(t)$, undergo a diffeomorphism, which mainly distorts the higher values in the signal -- the QRS complexes. Therefore, in this example, $\Omega_{f,\alpha}^c$ describes these QRS complexes and we expect the eigenvectors of operator $\mathbf{A}$ to be supported there. 
Figure \ref{fig:syntData} presents an example for the resulting simulated ta-mECG leads.

Using the simulated signals described above, we illustrate some of the properties of operators $\mathbf{S}$ and $\mathbf{A}$. 
We construct these operators according to Algorithm \ref{alg:SA}.
First, a lag-map is constructed from each signal, $s^{(\ell)}_i$, $\ell=1,2$,  in windows of $12$ samples and with an overlap of $6$ samples, in order to obtain a better representation of the data.
Denote the lag-map of signal $\ell$ by $s^{(\ell)}_{i,\ lag}$.
Second, an affinity matrix is constructed for each signal according to \eqref{eq:kernConst}, by treating each time frame (lag) as one sample, denoted by $x_i$ or $y_i$ in \eqref{eq:kernConst}.
The affinity matrices were constructed using the Euclidean distances between the samples, i.e. $d(x_i,x_j)=\left\Vert x_i-x_j\right\Vert_2$, and the the kernel scales, $\epsilon_1,\epsilon_2$, were set to be the median of the distances, which is common practice.
Third, operators $\mathbf{Q}^{(\ell)}$ and $\mathbf{P}^{(\ell)}$, $\ell=1,2$, are constructed for both $s^{(1)}_{i,\ lag}$ and $s^{(2)}_{i,\ lag}$ according to \eqref{eq:alg1_PQ}.
Finally, the operators $\mathbf{S}$ and $\mathbf{A}$ are constructed as in \eqref{eq:alg1_SA}.

\begin{figure}
\centering
\subfloat[]{\includegraphics[width=0.32\textwidth]{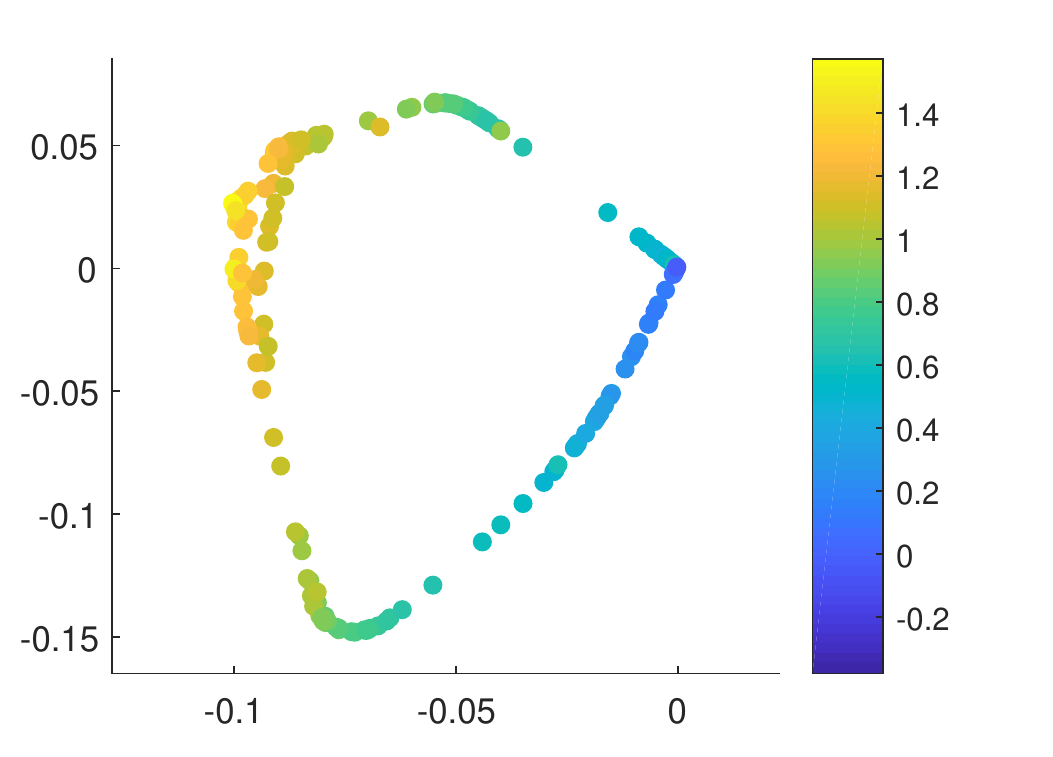}
}\subfloat[]{\includegraphics[width=0.32\textwidth]{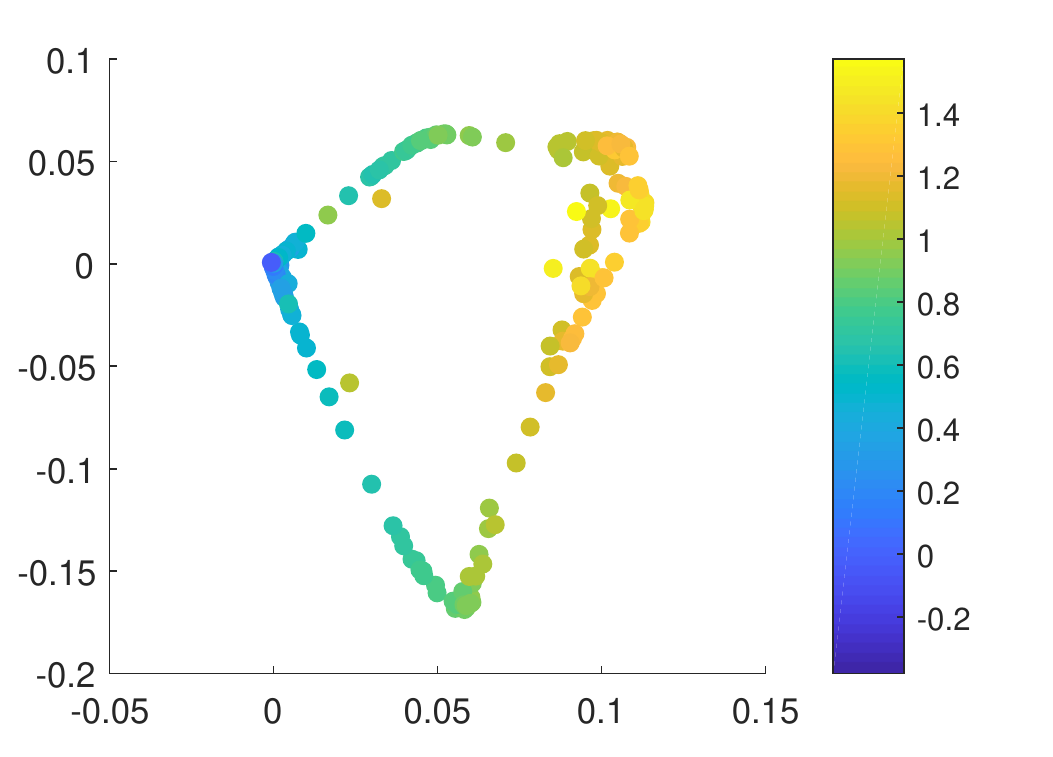}
}\subfloat[]{\includegraphics[width=0.32\textwidth]{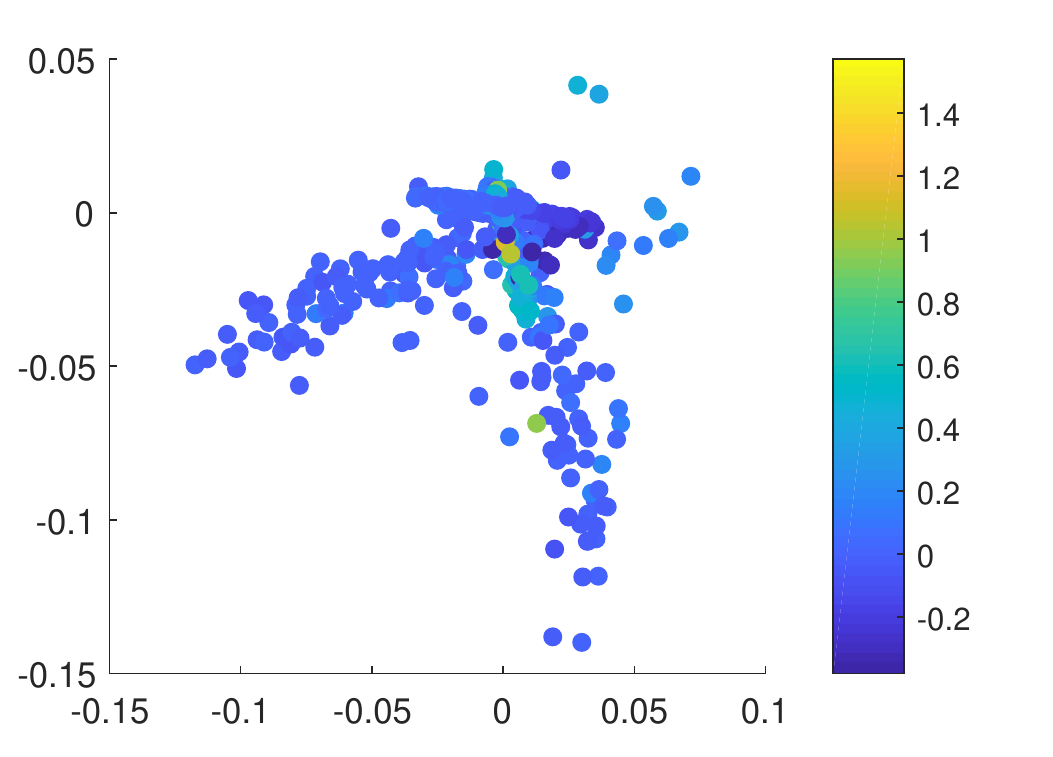}
}

\subfloat[]{\includegraphics[width=0.32\textwidth]{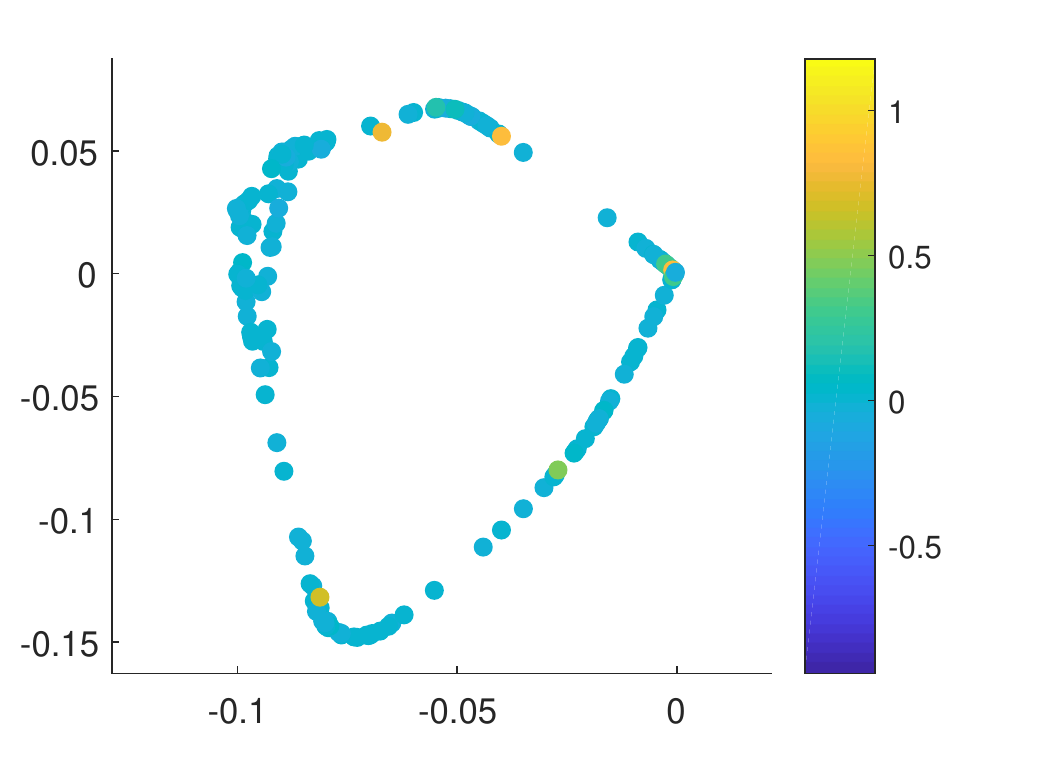}
}\subfloat[]{\includegraphics[width=0.32\textwidth]{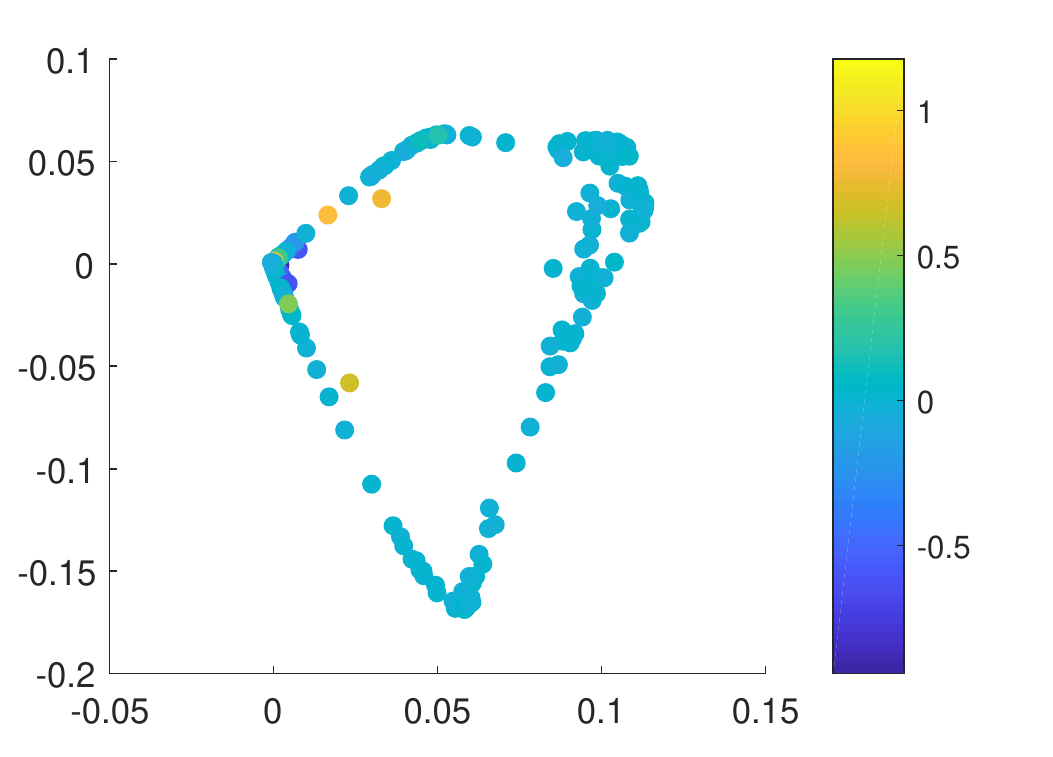}
}\subfloat[]{\includegraphics[width=0.32\textwidth]{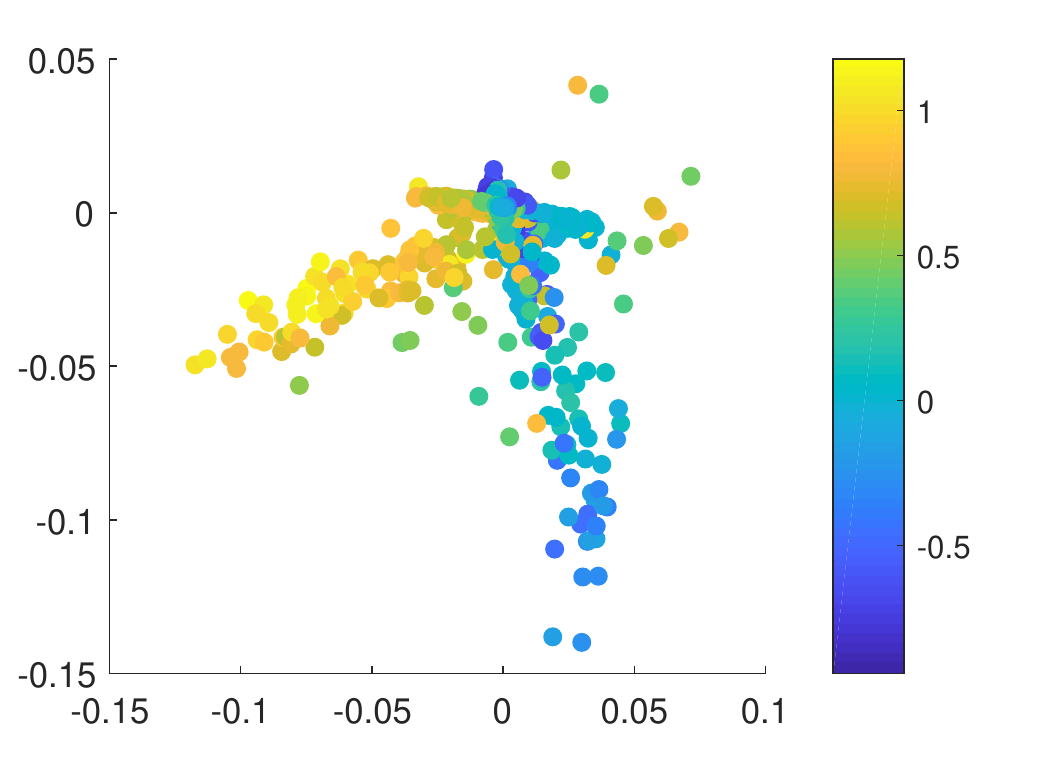}
}
\caption{Synthetic fetal heart rate detection example. Presenting the second and third eigenvectors of operator $\mathbf{S}$ (plots (b) and (e)) and the imaginary part of the second and third eigenvectors of operator $\mathbf{A}$ (plots (c) and (f)), compared with the first and second (non-trivial) eigenvectors of diffusion maps for ECG lead 1 (plots (a) and (d)). The plots are colored according to the maternal ECG in the upper row, and according to the fetal ECG in the bottom row.\label{fig:semiReal}}
\end{figure}

In Figure \ref{fig:semiReal}, scatter plots of the second and third eigenvectors of operators $\mathbf{S}$ and $\mathbf{A}$ are presented and compared to the eigenvectors of diffusion maps applied to each channel separately. Note that the choice to present the second and third eigenvectors of $\mathbf{S}$ and $\mathbf{A}$ is motivated by the result in Section \ref{sec:toyexamp}, where the respective first eigenvectors of $\mathbf{S}$ and $\mathbf{A}$ are similar and only represent the support of the non-isometric parts between the two manifolds.
In this figure, plots (a) and (d) depict $2$ eigenvectors (corresponding to the largest non-trivial eigenvalues) of diffusion maps, constructed based on the ECG lead $s^{(1)}_{i,\ lag}$. Plots (b) and (e) depict the second and third eigenvectors of the operator $\mathbf{S}$, and plots (c) and (f) depict the imaginary part of the second and third eigenvectors of the operator $\mathbf{A}$. The plots in the first row ((a), (b) and (c)) are colored according to the maternal ECG $z^{(1)}_i$, and the plots in the second row ((d), (e) and (f)) are colored according to one of the the fetal ECG signals $z^{(3)}_i$.

These plots show that in the eigenvectors of $\mathbf{A}$ the fetal ECG is significantly emphasized, compared with the eigenvectors of $\mathbf{S}$ and the diffusion maps embedding of the two channels. Furthermore, both the ECG lead $s^{(1)}_{i,\ lag}$ and the operator $\mathbf{S}$, which mainly describe the (more dominant) maternal ECG signal, lead to an embedding that corresponds to an embedding of $\mathcal{S}^1$, as can be seen in plots (a), (b), (d) and (e). This strengthens the model described in Subsection \ref{sub:fECGmodel}, in which the underlying manifolds $\mathcal{E}_m^{(1)}$ and $\mathcal{E}_m^{(2)}$ are diffeomorphic to $\mathcal{S}^1$. In contrast, the eigenvectors of $\mathbf{A}$ describe a different structure, since the difference between the ECG leads, $\Omega_{f,\alpha}^c$, is only a subset of $\mathcal{E}_f^{(1)}$ and $\mathcal{E}_f^{(2)}$.
We note that similar results were obtained for the second ECG lead, $s^{(2)}_{i,\ lag}$, and for the real part of the second and third eigenvectors of operator $\mathbf{A}$ and were omitted for brevity.

\begin{figure}
\centering
\subfloat[]{\includegraphics[width=0.9\textwidth]{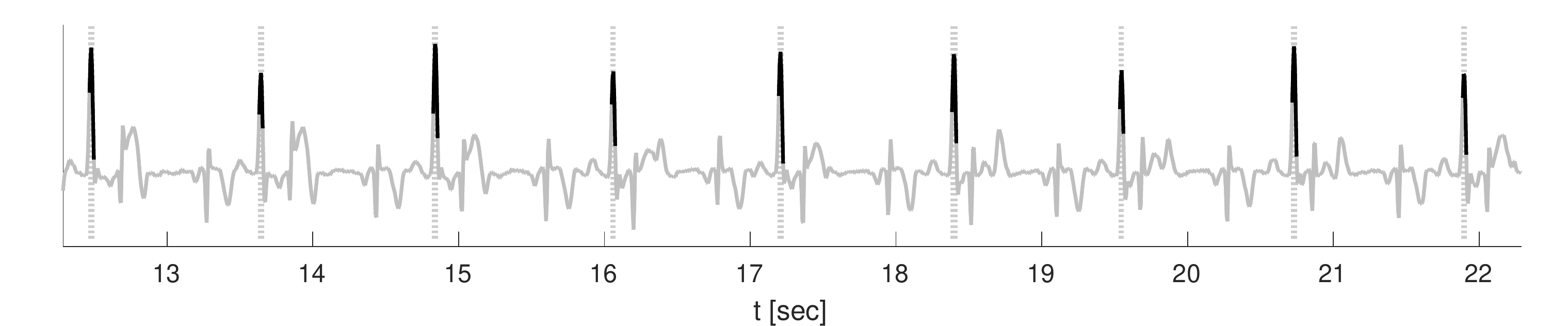}
}

\subfloat[]{\includegraphics[width=0.9\textwidth]{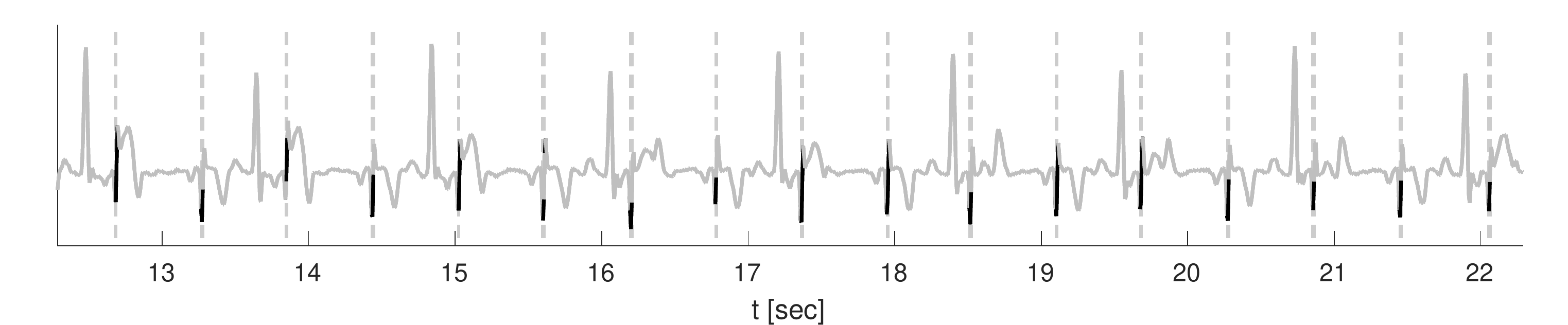}
}

\caption{ta-mECG segment from lead $s^{(1)}_i$, colored by (a) the second eigenvector of operator $\mathbf{S}$, (b) the imaginary part of the second eigenvector of operator $\mathbf{A}$. The vertical dotted lines in plot (a) mark the locations of the true maternal heart beats and the vertical dashed lines in plot (b) mark the locations of the true fetal heart beats.\label{fig:semiReal_ECG}}
\end{figure}

Figure \ref{fig:semiReal_ECG} presents a short simulated ta-mECG segment from lead $s^{(1)}_i$, containing both fetal and maternal components. Plots (a) and (b) are colored by an index vector, containing ones (colored in black) where the absolute value of the considered eigenvector exceeds a certain threshold and zeros (colored in gray) elsewhere. In plot (a), the segment is colored according to the second eigenvector of $\mathbf{S}$ with a threshold of $10^{-2}$, i.e. locations in which the eigenvector exceeds the threshold are colored in black. In plot (b), the segment is colored according to the imaginary part of the second eigenvector of $\mathbf{A}$ with a threshold of $3\times 10^{-2}$. The dotted vertical gray lines in plot (a) mark the locations of the true maternal beats and the dashed vertical gray lines in plot (b) mark the locations of the true fetal beats. These plots further demonstrate that $\mathbf{A}$ reveals the fetal beat locations, as the fetal heart beat morphologies are captured differently by the two synthetic leads.
In addition, the eigenvectors of $\mathbf{A}$ are supported mainly on the fetal QRS complexes, as assumed in the model presented in Subsection \ref{sub:fECGmodel}.

\subsection{Fetal heart rate detection -- real data\label{sub:realapp}}

Following the synthetic example in Subsection \ref{sub:semirealapp}, we address fHR detection from real ta-mECG recordings and propose to extract the fHR by constructing the operator $\mathbf{A}$ based on two ta-mECG leads. 
Similarly to the synthetic example, we expect that the operator $\mathbf{A}$ will provide a new representation of the signals which emphasizes the fetal beats.

We validate our approach using the publicly available database of ta-mECG signals, \href{https://physionet.org/challenge/2013/\#data-sets}{2013 PhysioNet/Computing in Cardiology Challenge}, abbreviated as CinC2013. We focus on the set A, which consists of 75 recordings, each of length 1 minute with R peak annotation and with reference to a ground-truth fECG signal, acquired from an invasive fetal scalp electrode. Each recording includes four noninvasive ta-mECG channels recorded from multiple positions using different electrodes (with possibly different configuration). The recordings are resampled at 1000 Hz. The lead placements on the maternal abdomen and the fetal/maternal health status are unknown. We disregard recording number 54 since it was excluded by the Challenge organizers \cite{andreotti2014robust}. In addition, we disregard recordings 33, 38, 47, 52, 71 and 74, since they contain inaccurate reference fetal annotations, as identified by \cite{behar2014combining}. We focus on the remaining 68 recordings.

\begin{figure}
\centering
\subfloat[]{\includegraphics[width=0.49\textwidth]{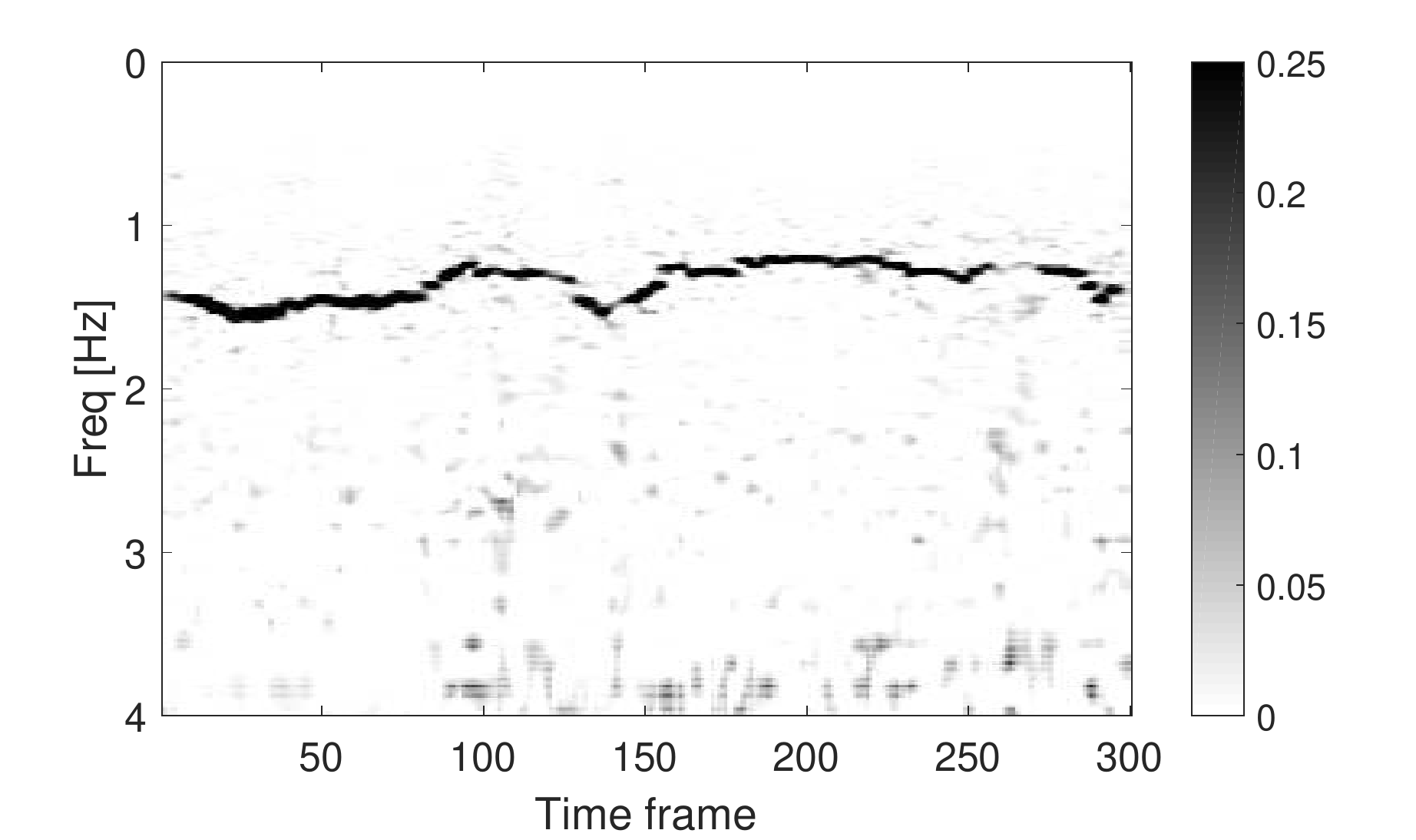}
}\subfloat[]{\includegraphics[width=0.49\textwidth]{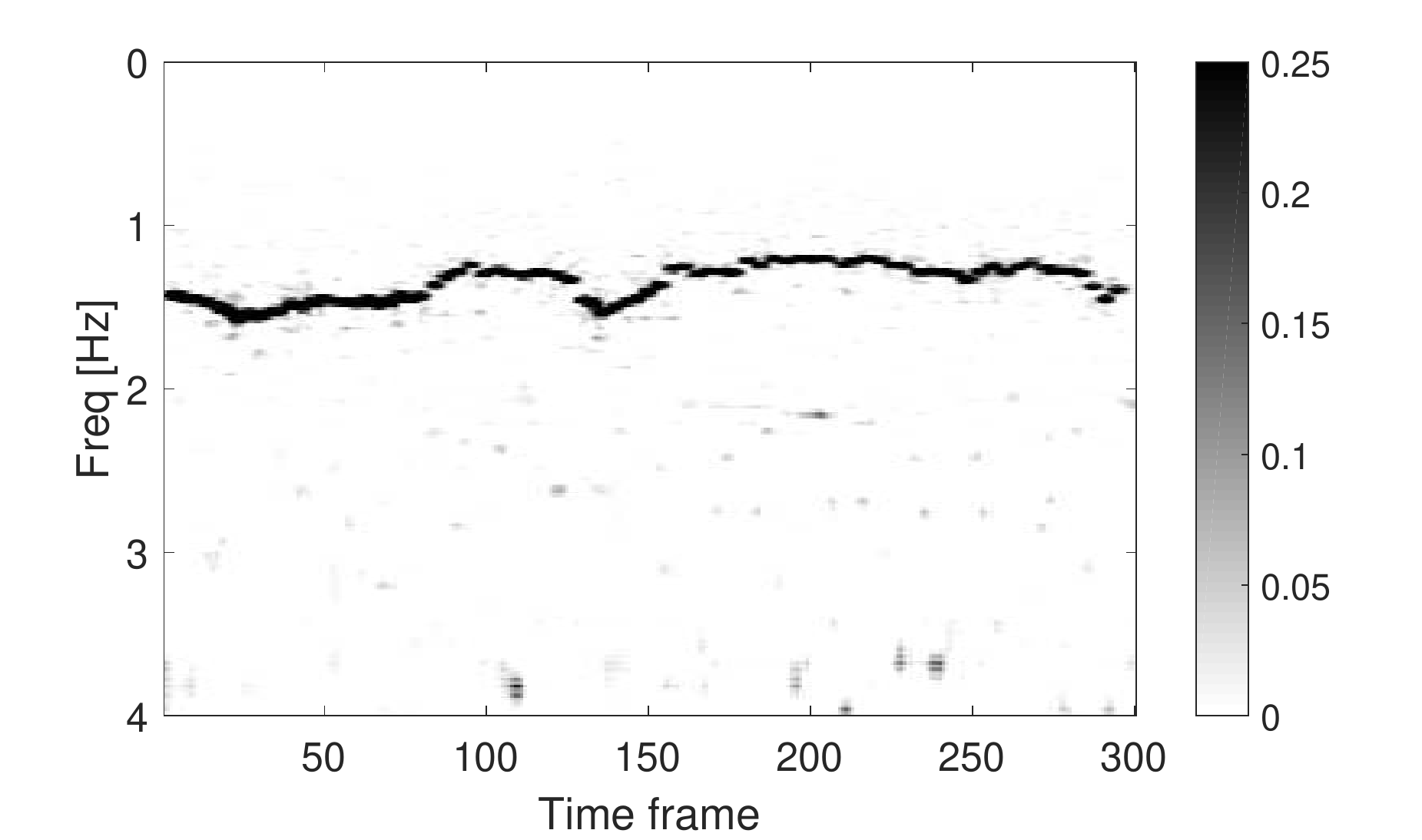}
}

\subfloat[]{\includegraphics[width=0.49\textwidth]{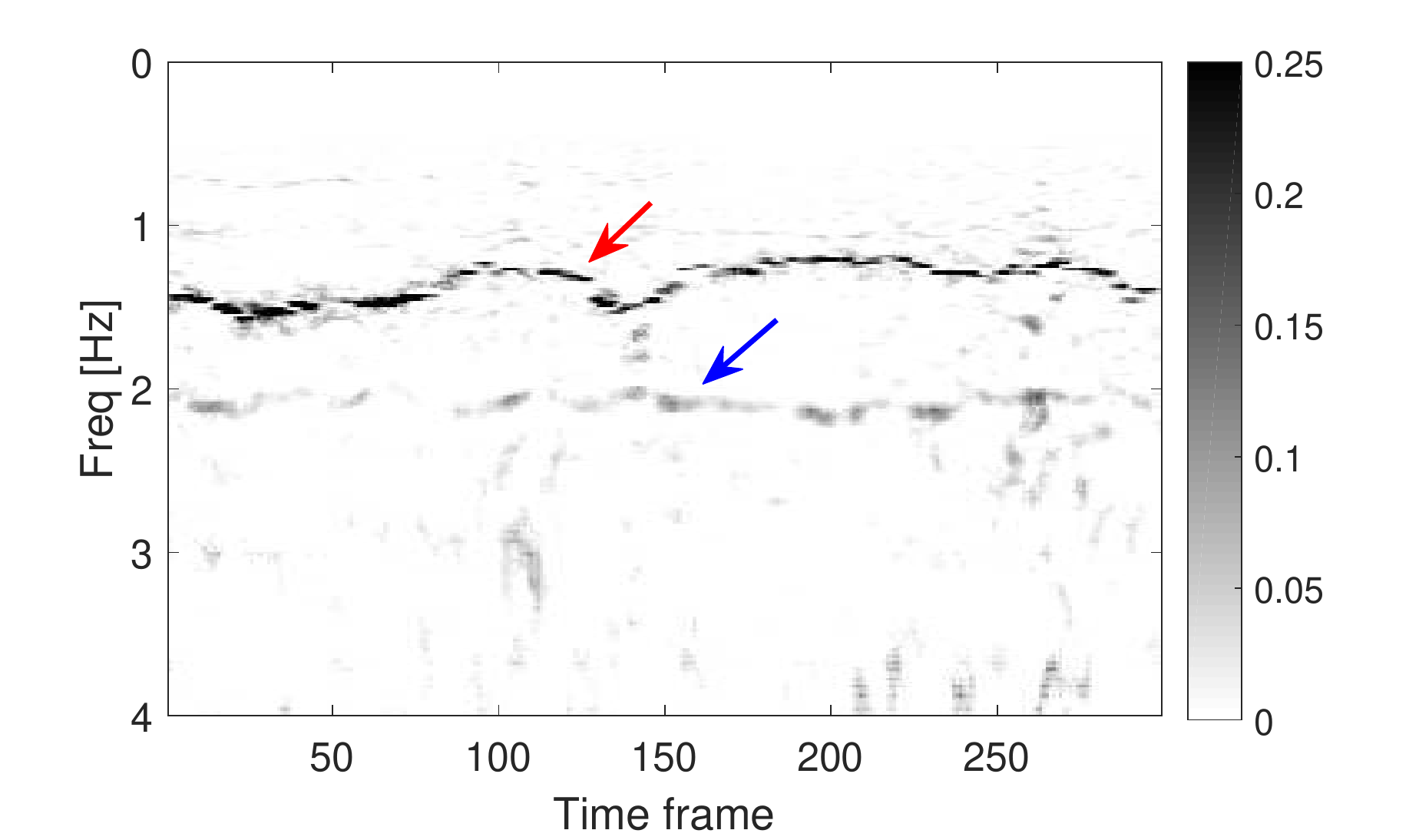}
}\subfloat[]{\includegraphics[width=0.49\textwidth]{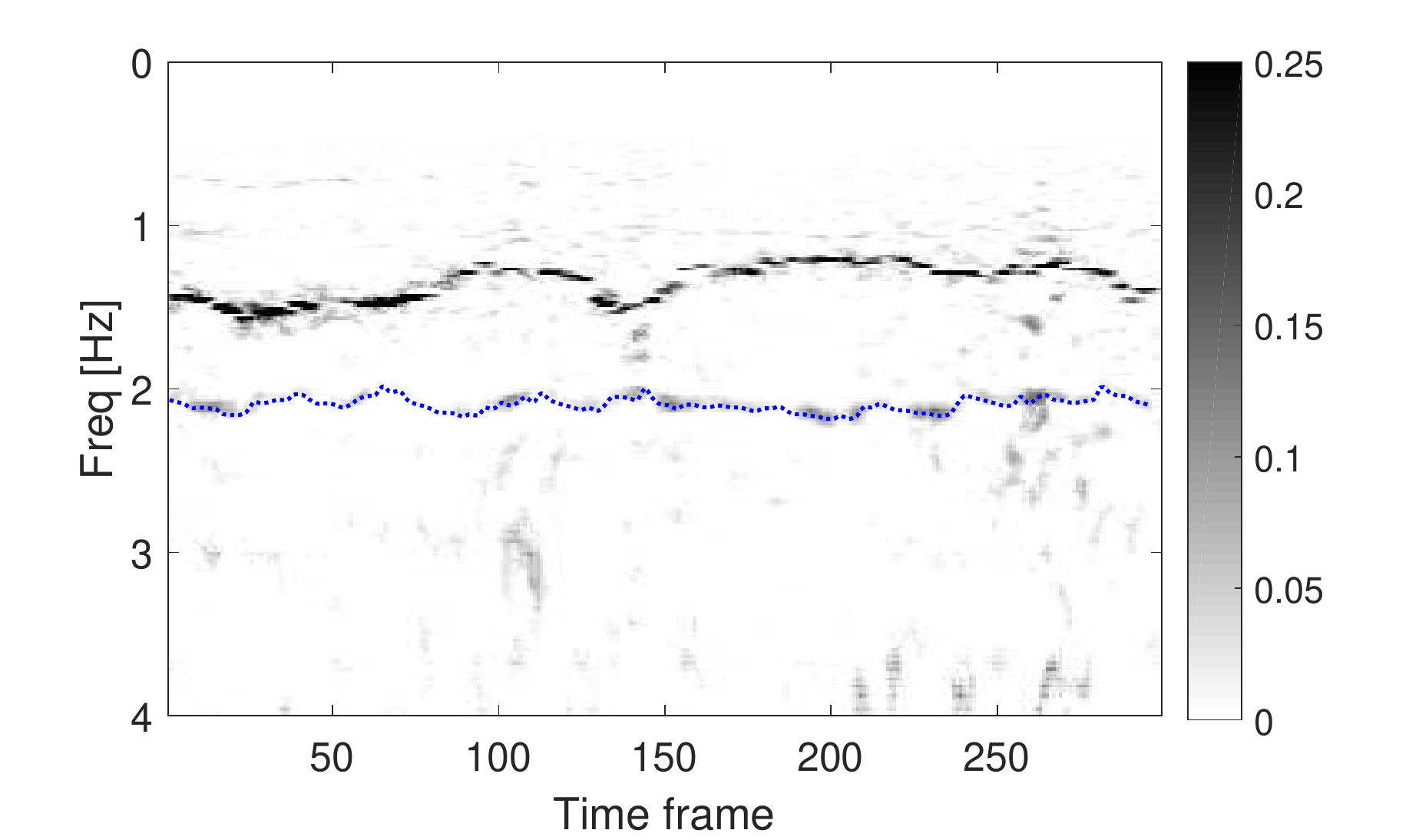}
}

\caption{Plots (a) and (b) present the spectrograms (after dsSTFT) of the ECG signal from two abdomen leads. The visually dominant frequency is the maternal ECG. Plot (c) presents the spectrogram extracted from the antisymmetric diffusion operator $\mathbf{A}$ (applied to the two abdomen signals). In this plot, the hidden fECG (marked by a blue arrow) is significantly enhanced.Plot (d) depicts the same spectrogram as plot (c), as well as the ground truth fetal heart rate (marked by a dotted blue line). \label{fig:fECGspect}}
\end{figure}

We first perform a pre-processing stage for each ta-ECG signal, which includes a low pass filter, below $100$[Hz], trend removal (median filtering with a window size of $101$ samples) and constructing a lag-map with a window of $8$ samples and a $7$-sample overlap.
After the pre-processing step, in the first stage of the proposed algorithm, we construct the forward and backward diffusion operators, $\mathbf{P}^{(\ell)},\mathbf{Q}^{(\ell)}$, $\ell=1,2$, from the lag-map of the two ta-mECG leads, and compute the operator $\mathbf{A}$ based on \eqref{eq:Aop}. 
We note that both forms of the anti-symmetric operator, $\mathbf{A}$ in \eqref{eq:Aop} and $\tilde{\mathbf{A}}$ in \eqref{eq:Atilde}, led to comparable results in this application (for both the synthetic and the real data). 
The eigenvectors of this operator are computed and sorted as described in Subsection \ref{sub:semirealapp}. 
The deshape Short Time Fourier Transform (dsSTFT) \cite{lin2016wave} is then applied to the real and imaginary parts of each of the first $20$ eigenvectors of $\mathbf{A}$, resulting in $40$ spectrograms, depicting the dominant frequencies in each eigenvector. 
The median (pixel-wise) over all of the dsSTFT spectrograms is taken as a new spectrogram for each subject, depicting both the fetal and maternal instantaneous heart rate. 
This can be viewed as a variation of the recently developed generalized multi-taper approach for time-frequency analysis, titled concentration of frequency and time \cite{daubechies2016conceft}.
Here, we use the eigenvectors of $\mathbf{A}$, which capture the oscillatory behavior of the signal, instead of the multiple windows in \cite{daubechies2016conceft}.
An example for such a spectrogram is presented in Figure \ref{fig:fECGspect}. 
In plots (a) and (b), the dsSTFT of the two ta-mECG leads are presented. 
The thick black line in these two plots represents the maternal heart rate. 
In plots (c), the median spectrogram of the eigenvectors of $\mathbf{A}$ is presented. 
In this plot, the red arrow marks the location of the maternal heart rate line and the blue arrow marks the location of the fetal heart rate line.
Plot (d) depicts the same spectrogram as plot (c) along with the ground truth of the fetal heart rate, marked by a dotted blue line.
Plots (c) and (d) demonstrate that the operator $\mathbf{A}$ leads to a result which significantly emphasizes the true fetal heart rate, compared with the original ta-mECG signals.

In the next stage of the algorithm, the fetal heart rate is extracted from the spectrogram, presented in Figure \ref{fig:fECGspect}(c). This is performed by first obtaining the maternal heart rate from the dsSTFT of the original ta-mECG signals (plots (a) and (b)) and removing its curve from the spectrogram of the operator $\mathbf{A}$. Second, the most dominant curve in the remaining spectrogram is extracted, using the algorithm described in \cite{li2017efficient}. This curve is assumed to represent the fetal heart rate. 
In order to extract the fetal ECG and the beat locations, we continue the analysis as described in \cite{li2017efficient}, after the dsSTFT stage.

The algorithm we applied to the ta-mECG leads is summarized in Algorithm \ref{alg:fECG}.
\begin{algorithm}
\caption{Fetal ECG Extraction Using Operator $\mathbf{A}$\label{alg:fECG}}
\begin{enumerate}
\item \textbf{Pre-processing} - Low pass filtering below $100$[Hz], trend removal (median filtering) and computation of a lag-map with a window of $8$ samples ($7$-sample overlap) for each ta-mECG lead. Denote the resulting signals by $s^{(\ell)}_{i,\ lag}$, $\ell=1,2$, where $i=1,\dots,N$ and $N=6\times 10^4$ is the number of samples in each ta-mECG recording.
\item \textbf{Fetal Instantaneous Heart Rate Detection}
\begin{enumerate}
\item Construct the operator $\mathbf{A}$ from the two ta-mECG leads, $s^{(1)}_{i,\ lag}$ and $s^{(2)}_{i,\ lag}$, and compute its eigenvectors $\psi_k(i)$ (sorted as in Subsection \ref{sub:semirealapp})\label{algst:Astage}.
\item Apply deshape Short Time Fourier Transform (dsSTFT) \cite{lin2016wave} to the real and imaginary parts of $\left\{\psi_k(i)\right\}_{k=1}^{20}$, separately.
\item Take the median over all the resulting spectrograms of the dsSTFT of $\left\{\psi_k(i)\right\}_{k=1}^{20}$.\label{algst:medSTFT}
\end{enumerate}
\item \textbf{Maternal ECG Removal and Fetal ECG Estimation} - Continue similarly to the algorithm described in \cite{li2017efficient}, after the dsSTFT stage, using the spectrogram obtained in step \ref{algst:medSTFT} above.
\end{enumerate}
\end{algorithm}

For performance evaluation we consider the F1 score, which is the harmonic mean of the sensitivity (SE) and the positive predictive value (PPV), similarly to \cite{li2017efficient}. The true positive (TP), false positive (FP) and false negative (FN) measures, used in the calculation of SE and PPV, were defined using a window of 50ms, i.e. a true positive classification means that an estimated beat is located within a window of 50ms around a true beat from the provided annotations.
We report the results of ta-mECG lead pair 1 and 4, which provided the best performance, out of the 6 possible pairs, for all algorithms. In addition, in the above performance measures, to avoid the boundary effect, the first and last 2 seconds in every recording are not evaluated. 

Table \ref{tab:F1fECGm} presents the F1 results obtained by the operators $\mathbf{A}$ and $\mathbf{S}$ using Algorithm \ref{alg:fECG}, as well as reference results obtained by the ta-mECG leads after the filtering in the pre-processing stage and application of PCA, denoted by $s_{PCA}^{(1)}$ and $s_{PCA}^{(2)}$. 
For the operator $\mathbf{A}$, Algorithm \ref{alg:fECG} is applied as is, whereas for operator $\mathbf{S}$, the only modification is the use of operator $\mathbf{S}$ instead of $\mathbf{A}$.
The mean, standard deviation (STD), median and interquartile range (IQR) over the F1 values of the $68$ subjects are presented. This table depicts that the operator $\mathbf{A}$ extracts significant information related to the fetal heart rate from the ta-mECG leads. It improves the results obtained by using the ta-mECG signals after applying PCA, $s_{PCA}^{(1)}$ and $s_{PCA}^{(2)}$.

\begin{table}
\centering
\begin{tabular}{| c || c | c | c | c |}
\hline 
& Mean(F1)\% & STD(F1)\% & median(F1)\% & IQR(F1)\% \\[0.5ex]
\hline\hline
$\mathbf{A}$ & $82.74$ & $28.37$ & $98.41$ & $12.7$\\[0.5ex]
\hline
$\mathbf{S}$ & $78.7$ & $28.66$ & $97.86$ & $46.31$\\[0.5ex]
\hline
$s_{PCA}^{(1)}$ & $73.01$ & $29.95$ & $94.13$ & $57.99$ \\[0.5ex]
\hline
$s_{PCA}^{(2)}$ & $78.59$ & $27.84$ & $97.02$ & $49.09$ \\[0.5ex]
\hline
\end{tabular}
\caption{F1 measure for the CinC2013 dataset calculated using Algorithm \ref{alg:fECG} with the operators $\mathbf{A}$ and $\mathbf{S}$ and compared to PCA applied directly to the ta-mECG signals\label{tab:F1fECGm}}
\end{table}

We note that the state-of-the-art results were obtained by \cite{li2017efficient}, which outperforms our results presented in Table \ref{tab:F1fECGm}. In \cite{li2017efficient}, the maternal ECG is first removed and then, only the remaining fetal ECG is processed. For comparison, we applied the proposed operators, $\mathbf{A}$ and $\mathbf{S}$, after first removing the maternal ECG in a similar manner. This led to improved results, which are closer to the state-of-the-art results. 
For $\mathbf{A}$, the median and IQR of the F1 measure in this case were $98.5\%$ and $6.7\%$ respectively, and for $\mathbf{S}$, they were $98.5\%$ and $6.2\%$ respectively. The mean and STD of the F1 measure were $87.3\% \pm 23.8\%$ for $\mathbf{A}$ and $87.1\% \pm 24.4\%$ for $\mathbf{S}$.
In this setting, the performance of operator $\mathbf{S}$ is comparable with operator $\mathbf{A}$, whereas in Table \ref{tab:F1fECGm}, the operator $\mathbf{A}$ leads to significantly better performance.
These results further demonstrate the properties of the proposed operators. In the latter case, removing the maternal ECG causes the fetal ECG to become the dominant {\it common} component in the two signals, which leads to its identification by the operator $\mathbf{S}$. 
In addition, since each ECG lead captures a different view of the fetal ECG it is still revealed by operator $\mathbf{A}$ as well.

While the reported performance does not outperform the state-of-the-art method based on two channels reported in \cite{li2017efficient}, these results support the potential of the anti-symmetric operator in fetal heart rate extraction, which was demonstrated by the synthetic example in Subsection \ref{sub:semirealapp} as well.

One of the reasons for the degraded performance of the operator $\mathbf{A}$ in the real application, compared with the synthetic example in Subsection \ref{sub:semirealapp}, is that the database is composed of heterogeneous signals -- it is recorded using different machines, includes pregnant women of different gestational ages, different lead placements, different noise levels, etc. (all of which are unknown to us).
In addition, the presence of significant noise in some ECG leads hampers the performance, since the noise is a part of the difference component (different between the two ECG leads) and  therefore, is captured by $\mathbf{A}$.

\section{Other related operators\label{sec:otherOps}}
The proposed operators are related to recent work, most of which concerning the recovery of common structures from different views, i.e. acquired by different modalities, or from different time frames, similarly to the symmetric operator $\mathbf{S}$. Such methods include the previously mentioned alternating diffusion \cite{lederman2015learning,talmon2016latent}, the dynamic Laplacian \cite{froyland2015dynamic,froyland2017dynamic}, cross-diffusion \cite{wang2012unsupervised} and the minimizing-disagreement algorithm \cite{de2005spectral}. 
One related work addressing the recovery of differences between shapes, similarly to $\mathbf{A}$, is presented in \cite{rustamov2013map}. 
In this section, we present a short overview of some of these related operators and discuss their connection to our work.

Most methods that address the recovery of common structures rely on operator composition. 
For example, the dynamic Laplacian \cite{froyland2015dynamic} focuses on recovering coherent sets in dynamical system, which can be modeled as the common structures in a set of manifolds, each representing a different time frame. The dynamic Laplacian operator is constructed from the composition of an operator with its adjoint, $\mathcal{L}^*_\epsilon\mathcal{L}_\epsilon$, where $\mathcal{L}_\epsilon=\mathit{P}_{2,\epsilon}\mathit{R}\mathit{P}_{1,\epsilon}$, and $\mathit{P}_{1,\epsilon}$ is a smoothing (diffusion) operator of the manifold corresponding to the first time frame, $\mathit{P}_{2,\epsilon}$ is a smoothing (diffusion) operator of the second time frame and $\mathit{R}$ is the Perron-Frobenius operator representing the system dynamics. In the context of our work, the operators $\mathit{P}_{1,\epsilon}$ and $\mathit{P}_{2,\epsilon}$ are analogous to the operator $P_\epsilon^{(\ell)}$.
In \cite{froyland2015dynamic} it is shown that this operator has a spectrum and converges to the sum of the Laplace-Beltrami operators of the manifolds representing the two time frames, similarly to $S_\epsilon$ as noted in Subsection \ref{sub:theoNew}. 
In the cross-diffusion algorithm \cite{wang2012unsupervised} two diffusion operators, $\mathbf{P}^{(1)}$ and $\mathbf{P}^{(2)}$, and their transpose, $\mathbf{Q}^{(1)}$ and $\mathbf{Q}^{(2)}$, are constructed (as in \eqref{eq:Pdisc} and \eqref{eq:Qdisc} respectively) based on two different metrics. A fusion of these metrics is then obtained by $\left[\mathbf{P}^{(1)}_{t+1}+\mathbf{P}^{(2)}_{t+1}\right]/2$, where $t>0$ and $\mathbf{P}^{(1)}_{t+1}:=\mathbf{P}^{(1)}\mathbf{P}^{(2)}_{t}\mathbf{Q}^{(1)}$ and $\mathbf{P}^{(2)}_{t+1}:=\mathbf{P}^{(2)}\mathbf{P}^{(1)}_{t}\mathbf{Q}^{(2)}$ are constructed iteratively. Note that, similarly to $\mathbf{S}$ and $\mathbf{A}$, this construction is also based on the composition of forward and backward operators, i.e. $\mathbf{Q}^{(\ell)}$ and $\mathbf{P}^{(\ell)}$ respectively.

Both of the above methods recover the common components only and ignore the differences. Therefore, compared with these operators, the novelty in the current work is the introduction of the difference revealing operator $\mathbf{A}$.
A similar notion of difference characterization between manifolds was previously presented in \cite{rustamov2013map}. There, a new linear operator for comparison of shape deformations was proposed, which provides a mapping between the shapes, and was shown to distort functions on the shapes only in areas where the shapes differ.
This operator was constructed as a composition of operators representing the two shapes, with one of the operators inverted, e.g. $(\mathbf{H}^{(1)})^{-1}\mathbf{F}^T\mathbf{H}^{(2)}\mathbf{F}$, where $\mathbf{H}^{(\ell)}$ denotes a matrix representing the inner product on shape $\ell$ and $\mathbf{F}$ is the functional map between the two shapes. In the context of our work, the operator $\mathbf{F}$ is analogous to the diffeomorphis, $\phi$, and the operator $\mathbf{H}^{(\ell)}$ is analogous to the operator $P^{(\ell)}_\epsilon$. In contrast to the proposed self-adjoint operator $\mathbf{A}$, this shape difference operator does not necessarily have a spectrum and depends on the order of the operator composition. In addition, in the discrete setting, it requires the inverse (or pseudo inverse) of a possibly large matrix.

Other operators for recovering differences between manifolds can be considered. For example, $\mathbf{\hat{A}}=\left(\mathbf{P}^{(1)}-\mathbf{P}^{(2)}\right) \left(\mathbf{P}^{(1)}-\mathbf{P}^{(2)}\right)^T$ is a symmetric operator which obtains comparable results in the experimental results in Section \ref{sec:toyexamp} and Section \ref{sec:fECG}. However, this operator was not considered in the current paper since in the asymptotic expansion of this operator, in contrast to $A_\epsilon$, the second order terms, of order $O(\epsilon^2)$, cancel out and only fourth order terms and above remain.
In the future, we plan to extend this work and explore such additional operators for recovering hidden components of multimodal data, and create a ``library'' of operators. We plan to use this library of operators and construct a framework for characterizing the common and the difference structures in multimodal data or in data which lies on a time evolving manifold. Moreover, we plan to devise a multi-resolution analysis framework for time-varying manifolds based on such a library of operators, which can be seen as analogous to the wavelet analysis under the manifold setting.

Note the assumption hidden in both the composite operators $\mathbf{S}$ and $\mathbf{A}$ and in the presented alternative operator, $\mathbf{\hat{A}}$. The addition and subtraction operations in the composition imply that the operators lie in a linear Euclidean space, which may violate the Riemannian structure of the operators. In future work, we plan to address this issue and investigate different ways of composing such operators using non-Euclidean settings.

\begin{appendices}

\section{Proof of Proposition \ref{prop:PQ} for the operator $Q_\epsilon$\label{app:propPQ}}
In this appendix we show that the asymptotic expansion of the operator $Q_\epsilon$, presented in Subsection \ref{sub:TheoSingle}, is given by
\begin{equation}
Q_{\epsilon}f\left(x\right) = \int k_\epsilon(x,x')\frac{f(x')\mu(x')}{\hat{d}_\epsilon(x')}dV(x') = f(x) - \epsilon^2 \left( \Delta f(x) - \frac{f\Delta\mu}{\mu}(x) \right) + O(\epsilon^4),
\end{equation}
where $\hat{d}_\epsilon(x')=\int k_\epsilon(x,x')\mu(x)dV(x)$.

\begin{proof}
As shown in \cite{Coifman2006} (Appendix B, Lemma 8), the asymptotic expansion of an appropriately scaled kernel $k_\epsilon(x,x')$, defined similarly to \eqref{eq:dmkern}, applied to any smooth function $g(x)$ on $\mathcal{M}$, is given by
\begin{equation}
K_\epsilon g(x) = \int k_\epsilon(x,x')g(x')dV(x') = g(x) - \epsilon^2\left(\Delta g(x) - \omega(x)g(x)\right) + O(\epsilon^4),
\end{equation}
where $\omega(x)$ is a function that depends on the curvature.

Therefore, for $Q_{\epsilon}$, consider $g(x)=f(x)\mu(x)/\hat{d}_\epsilon(x)$, and its asymptotic expansion is given by
\begin{equation}
Q_{\epsilon}f\left(x\right) = \frac{f(x)\mu(x)}{\hat{d}_\epsilon(x)} - \epsilon^2\left(\Delta\left(\frac{f\mu}{\hat{d}_\epsilon}\right)(x) - \omega(x)\frac{f(x)\mu(x)}{\hat{d}_\epsilon(x)}\right) + O(\epsilon^4).\label{eq:Qderiv1}
\end{equation}

\begin{sloppypar}
In addition, for $\hat{d}_\epsilon(x)$, consider $g(x)=\mu(x)$ and then $\hat{d}_\epsilon(x)=\mu(x) - \epsilon^2\left(\Delta\mu(x) - \omega(x)\mu(x)\right) + O(\epsilon^4)$. When $\epsilon$ is sufficiently small, we have, 
\begin{equation}
\left(\hat{d}_\epsilon\right)^{-1}=\left(\mu\right)^{-1}\left(1 + \epsilon^2\left(\frac{\Delta\mu}{\mu}-\omega\right)\right)+O(\epsilon^4).\label{eq:hatwDeriv}
\end{equation}
\end{sloppypar}

By substituting $\hat{d}_\epsilon$ in \eqref{eq:Qderiv1} with \eqref{eq:hatwDeriv}, when $\epsilon$ is sufficiently small, we obtain the following asymptotic expansion
\begin{eqnarray}
Q_{\epsilon}f\left(x\right) & = & f(x)-\epsilon^2\left(\Delta f(x)  - \omega(x)f(x) + \omega(x)f(x) - f\frac{\Delta\mu}{\mu}(x)\right) + O(\epsilon^4)\\
& = & f(x)-\epsilon^2\left(\Delta f(x) - f\frac{\Delta\mu}{\mu}(x)\right) + O(\epsilon^4).
\end{eqnarray}
\end{proof}

\section{Proof of Proposition \ref{prop:GH}\label{app:propGH}}
For simplicity, we present the proof of Proposition \ref{prop:GH} for $\epsilon_2=\epsilon_1=\epsilon$. For $\epsilon_2\neq\epsilon_1$,  the proof is similar up to some notation changes. The asymptotic expansion of the operators $G_\epsilon$ and $H_\epsilon$, defined in Subsection \ref{sub:theoAd}, is given by
\begin{align}
G_{\epsilon} f(x) = f(x)&  - \epsilon^2\left(\Delta ^{(1)}f(x) + \phi^*\Delta^{(2)}(\phi^*)^{-1}f(x)\right)\\
& - \epsilon^2\left(\phi^*\frac{2\nabla^{(2)}(\phi^*)^{-1}f\cdot\nabla^{(2)}\mu^{(2)}}{\mu^{(2)}}(x) - \frac{f\Delta^{(1)}\mu^{(1)}} {\mu^{(1)}}(x)\right) + O(\epsilon^4)\\
H_\epsilon f(x) = f(x) &- \epsilon^2\left(\phi^*\Delta^{(2)}(\phi^*)^{-1}f(x) + \Delta ^{(1)}f(x)\right)\\
& - \epsilon^2\left(\frac{2\nabla^{(1)}f\cdot\nabla^{(1)}\mu^{(1)}}{\mu^{(1)}}(x) - f\phi^*\frac{\Delta^{(2)}\mu^{(2)}} {\mu^{(2)}}(x)\right) + O(\epsilon^4)
\end{align}

\begin{proof}
From Proposition \ref{prop:PQ}, for $x\in\mathcal{M}^{(\ell)}$, we have
\begin{align}
P_{\epsilon}^{(\ell)}f\left(x\right) &= f(x) - \epsilon^2 \left( \Delta ^{(\ell)} f + \frac{ 2\nabla^{(\ell)}f\cdot\nabla^{(\ell)}\mu^{(\ell)}}{\mu^{(\ell)}} \right) (x) + O(\epsilon^4) \\
Q_{\epsilon}^{(\ell)}f\left(x\right) &= f(x) - \epsilon^2\left( \Delta^{(\ell)}f - \frac{f\Delta^{(\ell)}\mu^{(\ell)}}{\mu^{(\ell)}} \right) (x) + O(\epsilon^4).
\end{align}

\begin{sloppypar}
For the operator $G_\epsilon f(x)=\phi^* P_\epsilon^{(2)} (\phi^*)^{-1} Q_\epsilon^{(1)}f(x)$, where $x\in\mathcal{M}^{(1)}$, consider $g(y)=\left((\phi^*)^{-1}Q_\epsilon^{(1)}f\right)(y)$, where $y=\phi(x)$, and place the expansion of $\left((\phi^*)^{-1}Q_\epsilon^{(1)}f\right)(y)$ into $\left(\phi^*P_{\epsilon}^{(2)}g\right)(x)$:
\begin{align}
G_\epsilon f(x) = & \left(\phi^*P_{\epsilon}^{(2)}g\right)(x)\\
= & \phi^*\left[g-\epsilon^2\left(\Delta^{(2)}g + \frac{2\nabla^{(2)}g\cdot\nabla^{(2)}\mu^{(2)}}{\mu^{(2)}}\right)\right](x) + O(\epsilon^4) \\
= & f(x) - \epsilon^2\left(\Delta^{(1)}f - \frac{f\Delta^{(1)}\mu^{(1)}}{\mu^{(1)}}\right)(x) \\
&  - \epsilon^2\left(\phi^*\Delta^{(2)}(\phi^*)^{-1}f + \phi^*\frac{2\nabla^{(2)}(\phi^*)^{-1}f\cdot\nabla^{(2)}\mu^{(2)}}{\mu^{(2)}}\right)(x) + O(\epsilon^4).
\end{align}
\end{sloppypar}

Similarly, for $H_\epsilon$ we get
\begin{eqnarray}
H_\epsilon f(x) = & f(x)& - \epsilon^2\left(\Delta ^{(1)}f(x) + \phi^*\Delta^{(2)}(\phi^*)^{-1}f(x)\right)\\
& & - \epsilon^2\left(\frac{2\nabla^{(1)}f\cdot\nabla^{(1)}\mu^{(1)}}{\mu^{(1)}}(x) - f\phi^*\frac{\Delta^{(2)}\mu^{(2)}} {\mu^{(2)}}(x)\right) + O(\epsilon^4).
\end{eqnarray}
\end{proof}

\begin{remark}
The difference between the asymptotic expansions of the operators $G_\epsilon$ and $H_\epsilon$ and the alternating diffusion operator shown in Appendix \ref{app:compareAD}, is in the term $f\frac{\Delta^{(\ell)}\mu^{(\ell)}} {\mu^{(\ell)}}$, which appears in $G_\epsilon$ and $H_\epsilon$. In the alternating diffusion operator the expressions representing the two manifolds are similar and given by $\frac{2\nabla^{(\ell)}f\cdot\nabla^{(\ell)}\mu^{(\ell)}}{\mu^{(\ell)}}$.
\end{remark}

\section{Proof of Proposition \ref{prop:SA}\label{app:propSA}}
For simplicity, we present the proof of Proposition \ref{prop:SA} for $\epsilon_2=\epsilon_1=\epsilon$. For $\epsilon_2\neq\epsilon_1$, the proof is similar up to some notation changes. For the operators $S_\epsilon$ and $A_\epsilon$, defined in Subsection \ref{sub:theoNew}, we present the derivation of the asymptotic expansion and prove Proposition \ref{prop:SA}.

\begin{proof}
For $S_\epsilon f(x)$, place the asymptotic expansions of $G_\epsilon$ and $H_\epsilon$, shown in Proposition \ref{prop:GH}, into $S_\epsilon f(x) = (G_\epsilon f(x) + H_\epsilon f(x))/2$ to obtain:
\begin{align}
S_\epsilon f(x) = &\frac{1}{2}f(x) - \frac{\epsilon^2}{2}\left(\Delta^{(1)}f(x) + \phi^*\Delta^{(2)}(\phi^*)^{-1}f(x)\right)\\
&  - \frac{\epsilon^2}{2}\left(\phi^*\frac{2\nabla^{(2)}(\phi^*)^{-1}f\cdot\nabla^{(2)}\mu^{(2)}}{\mu^{(2)}}(x) - \frac{f\Delta^{(1)}\mu^{(1)}}{\mu^{(1)}}(x)\right)\\
& + \frac{1}{2}f(x) - \frac{\epsilon^2}{2}\left(\phi^*\Delta^{(2)}(\phi^*)^{-1}f(x) + \Delta^{(1)}f(x)\right)\\
&  - \frac{\epsilon^2}{2}\left(\frac{2\nabla^{(1)}f\cdot\nabla^{(1)}\mu^{(1)}}{\mu^{(1)}}(x) - f\phi^*\frac{\Delta^{(2)}\mu^{(2)}} {\mu^{(2)}}(x)\right) + O(\epsilon^4)\\
= &f(x) - \epsilon^2\left(\Delta^{(1)}f(x) + \phi^*\Delta^{(2)}(\phi^*)^{-1}f(x)\right)\\
&  - \frac{\epsilon^2}{2}\left(\phi^*\frac{2\nabla^{(2)}(\phi^*)^{-1}f\cdot\nabla^{(2)}\mu^{(2)}}{\mu^{(2)}}(x) - f\phi^*\frac{\Delta^{(2)}\mu^{(2)}}{\mu^{(2)}}(x)\right)\\
&  - \frac{\epsilon^2}{2}\left(\frac{2\nabla^{(1)}f\cdot\nabla^{(1)}\mu^{(1)}}{\mu^{(1)}}(x) - \frac{f\Delta^{(1)}\mu^{(1)}}{\mu^{(1)}}(x)\right) + O(\epsilon^4).
\end{align}

For $A_\epsilon f(x)$, place the asymptotic expansions of $G_\epsilon$ and $H_\epsilon$, shown in Proposition \ref{prop:GH}, into $A_\epsilon f(x) = (G_\epsilon f(x) - H_\epsilon f(x))/2$ to obtain:
\begin{align}
A_\epsilon f(x) = &\frac{1}{2}f(x) - \frac{\epsilon^2}{2}\left(\Delta^{(1)}f(x) + \phi^*\Delta^{(2)}(\phi^*)^{-1}f(x)\right)\\
&  - \frac{\epsilon^2}{2}\left(\phi^*\frac{2\nabla^{(2)}(\phi^*)^{-1}f\cdot\nabla^{(2)}\mu^{(2)}}{\mu^{(2)}}(x) - \frac{f\Delta^{(1)}\mu^{(1)}}{\mu^{(1)}}(x)\right)\\
& - \frac{1}{2}f(x) + \frac{\epsilon^2}{2}\left(\phi^*\Delta^{(2)}(\phi^*)^{-1}f(x) + \Delta^{(1)}f(x)\right)\\
&  + \frac{\epsilon^2}{2}\left(\frac{2\nabla^{(1)}f\cdot\nabla^{(1)}\mu^{(1)}}{\mu^{(1)}}(x) - f\phi^*\frac{\Delta^{(2)}\mu^{(2)}} {\mu^{(2)}}(x)\right) + O(\epsilon^4)\\
= &  \frac{\epsilon^2}{2}\left(\frac{2\nabla^{(1)}f\cdot\nabla^{(1)}\mu^{(1)}}{\mu^{(1)}}(x) + \frac{f\Delta^{(1)}\mu^{(1)}}{\mu^{(1)}}(x)\right)\\
&  - \frac{\epsilon^2}{2}\left(\phi^*\frac{2\nabla^{(2)}(\phi^*)^{-1}f\cdot\nabla^{(2)}\mu^{(2)}}{\mu^{(2)}}(x) + f\phi^*\frac{\Delta^{(2)}\mu^{(2)}} {\mu^{(2)}}(x)\right) + O(\epsilon^4).
\end{align}
\end{proof}

\section{Comparison to alternating diffusion\label{app:compareAD}}
In this appendix, we review the asymptotic expansion of the alternating diffusion operator from \cite{talmon2016latent,lederman2015learning} and show that it is not self-adjoint. For simplicity, we assume that $\epsilon_2=\epsilon_1=\epsilon$. For $\epsilon_2\neq\epsilon_1$, the derivations are similar up to some notation changes.

The asymptotic expansion of the alternating diffusion operator can be derived similarly to Appendix \ref{app:propGH} and Appendix \ref{app:propSA}. This operator is defined by $P^{AD}_\epsilon f(x)=\phi^* P_\epsilon^{(2)}(\phi^*)^{-1} P_\epsilon^{(1)}f(x)$. By placing the asymptotic expansion of $P_\epsilon^{(\ell)}$ from Proposition \ref{prop:PQ} in this definition we get
\begin{align}
P^{AD}_\epsilon f(x) & = f(x) - \epsilon^2\left(\Delta^{(1)}f + \frac{2\nabla^{(1)}f\cdot\nabla^{(1)}\mu^{(1)}}{\mu^{(1)}}\right)(x)\label{eq:ad_m1}\\
& - \epsilon^2\left(\phi^*\Delta^{(2)}(\phi^*)^{-1} f + \phi^*\frac{2\nabla^{(2)}(\phi^*)^{-1} f\cdot\nabla^{(2)}\mu^{(2)}}{\mu^{(2)}}\right)(x) + O(\epsilon^4)\label{eq:ad_m2}\\
& = f(x) - \epsilon^2\left(\Delta^{(1)}f + \phi^*\Delta^{(2)}(\phi^*)^{-1} f\right)(x)\\
& - \epsilon^2\left(\frac{2\nabla^{(1)}f\cdot\nabla^{(1)}\mu^{(1)}}{\mu^{(1)}} + \phi^*\frac{2\nabla^{(2)}(\phi^*)^{-1} f\cdot\nabla^{(2)}\mu^{(2)}}{\mu^{(2)}}\right)(x) + O(\epsilon^4).
\end{align}

\begin{sloppypar}
We now show that the limit operator of alternating diffusion, $P^{AD}=\lim_{\epsilon\rightarrow 0}\left(I - P^{AD}_\epsilon\right)/\epsilon^2$, where $I$ denotes the identity operator, is not self-adjoint. 
We separate $P^{AD}$ into two additive terms, the first, denoted by $P^{AD(1)}$, which contains elements related to the first manifold, i.e. elements from \eqref{eq:ad_m1}, and the second, denoted by $P^{AD(2)}$, which contains elements related to the second manifold, i.e. elements from \eqref{eq:ad_m2}. We will show that each of these operators is not self-adjoint, and therefore, $P^{AD}$ is not self-adjoint, from the linearity of the inner product and from the additivity of these operators.
\end{sloppypar}

For $P^{AD(1)}$, given $f,g\in C^\infty\left(\mathcal{M}^{(1)}\right)$,
\begin{align}
\left\langle P^{AD(1)}f,g \right\rangle_{\mathcal{M}^{(1)}} & = \int_{\mathcal{M}^{(1)}}\left(\Delta^{(1)}f + \frac{2\nabla^{(1)}f\cdot\nabla^{(1)}\mu^{(1)}}{\mu^{(1)}}\right)(x)g(x)\mu^{(1)}(x)dV^{(1)}(x)\nonumber\\
& = \int_{\mathcal{M}^{(1)}}\left(\Delta^{(1)}f(x)\right)g(x)\mu^{(1)}(x)dV^{(1)}(x)\nonumber\\
& + \int_{\mathcal{M}^{(1)}}\left(2\nabla^{(1)}f\cdot\nabla^{(1)}\mu^{(1)}\right)(x)g(x)dV^{(1)}(x)\label{eq:ad_self_adj_preGreen}\\
& = \int_{\mathcal{M}^{(1)}}\left(\Delta^{(1)}g+\frac{2\nabla^{(1)}g\cdot\nabla^{(1)}\mu^{(1)}}{\mu^{(1)}}\right)(x)\mu^{(1)}(x)f(x)dV^{(1)}(x)\nonumber\\ 
& + \int_{\mathcal{M}^{(1)}}\left(g\frac{\Delta^{(1)}\mu^{(1)}}{\mu^{(1)}}\right)(x)\mu^{(1)}(x)f(x)dV^{(1)}(x)\nonumber\\
& - \int_{\mathcal{M}^{(1)}}\left(\frac{2\nabla^{(1)}g\cdot\nabla^{(1)}\mu^{(1)}}{\mu^{(1)}}+2g\frac{\Delta^{(1)}\mu^{(1)}}{\mu^{(1)}}\right)(x)\mu^{(1)}(x)f(x)dV^{(1)}(x)\label{eq:ad_self_adj_postGreen1}\\
& = \int_{\mathcal{M}^{(1)}}\left(\Delta^{(1)}g - g\frac{\Delta^{(1)}\mu^{(1)}}{\mu^{(1)}}\right)(x)\mu^{(1)}(x)f(x)dV^{(1)}(x)\\
& \neq \left\langle f,P^{AD(1)}g \right\rangle_{\mathcal{M}^{(1)}},
\end{align}
where the transition between \eqref{eq:ad_self_adj_preGreen} and \eqref{eq:ad_self_adj_postGreen1}, is based on Green's first identity (for manifolds without a boundary).

Similarly, for $P^{AD(2)}$, given $f,g\in C^\infty\left(\mathcal{M}^{(1)}\right)$,
\begin{eqnarray}
\left\langle P^{AD(2)}f,g \right\rangle_{\mathcal{M}^{(1)}} & = & \int_{\mathcal{M}^{(1)}}\left(\phi^*\Delta^{(2)}(\phi^*)^{-1}f\right)(x)g(x)\mu^{(1)}(x)dV^{(1)}(x)\nonumber\\ 
& & + \int_{\mathcal{M}^{(1)}}\phi^*\frac{2\nabla^{(2)}(\phi^*)^{-1} f\cdot\nabla^{(2)}\mu^{(2)}}{\mu^{(2)}}(x)g(x)\mu^{(1)}(x)dV^{(1)}(x)\label{eq:ad_self_adj1}\\
& = & \int_{\mathcal{M}^{(2)}}\left((\phi^*)^{-1}g\mu^{(2)}\Delta^{(2)}(\phi^*)^{-1}\right)(y)f(y)dV^{(2)}(y)\nonumber\\
& & + \int_{\mathcal{M}^{(2)}}\left(2(\phi^*)^{-1}g\nabla^{(2)}(\phi^*)^{-1}f\cdot\nabla^{(2)}\mu^{(2)}\right)(y)dV^{(2)}(y)\label{eq:ad_self_adj2}\\
& = & \int_{\mathcal{M}^{(2)}}\left(\mu^{(2)}(\phi^*)^{-1}f\Delta^{(2)}(\phi^*)^{-1}g\right)(y)dV^{(2)}(y)\nonumber\\
&& + \int_{\mathcal{M}^{(2)}}\left(2(\phi^*)^{-1}f\nabla^{(2)}(\phi^*)^{-1}g\cdot\nabla^{(2)}\mu^{(2)}\right)(y)dV^{(2)}(y)\nonumber\\
&& +\int_{\mathcal{M}^{(2)}}\left((\phi^*)^{-1}f(\phi^*)^{-1}g\Delta^{(2)}\mu^{(2)}\right)(y)dV^{(2)}(y)\nonumber\\ 
& & - \int_{\mathcal{M}^{(2)}}\left(2(\phi^*)^{-1}f\nabla^{(2)}(\phi^*)^{-1}g\cdot\nabla^{(2)}\mu^{(2)}\right)(y)dV^{(2)}(y)\nonumber\\
& & + \int_{\mathcal{M}^{(2)}}\left(2(\phi^*)^{-1}f(\phi^*)^{-1}g\Delta^{(2)}\mu^{(2)}\right)(y)dV^{(2)}(y)\label{eq:ad_self_adj3}\\
& = & \int_{\mathcal{M}^{(2)}}\left((\phi^*)^{-1}f\Delta^{(2)}(\phi^*)^{-1}g\right)(y)\mu^{(2)}(y)dV^{(2)}(y)\nonumber\\
&& - \int_{\mathcal{M}^{(2)}}\left((\phi^*)^{-1}f(\phi^*)^{-1}g\frac{\Delta^{(2)}\mu^{(2)}}{\mu^{(2)}}\right)(y)\mu^{(2)}(y)dV^{(2)}(y)\label{eq:ad_self_adj4}\\
& = & \int_{\mathcal{M}^{(1)}}\left(\phi^*\Delta^{(2)}(\phi^*)^{-1}g - g\phi^*\frac{\Delta^{(2)}\mu^{(2)}}{\mu^{(2)}}\right)(x)f(x)\mu^{(1)}(x)dV^{(1)}(x)\label{eq:ad_self_adj5}\\
 & \neq & \left\langle f,P^{AD(2)}g \right\rangle_{\mathcal{M}^{(1)}},
\end{eqnarray}
where the transitions from \eqref{eq:ad_self_adj1} to \eqref{eq:ad_self_adj2} and from \eqref{eq:ad_self_adj4} to \eqref{eq:ad_self_adj5} are based on $\mu^{(1)}(x)dV^{(1)}(x)=\mu^{(2)}(y)dV^{(2)}(y)$ and $y=\phi(x)$. In addition, the transition between \eqref{eq:ad_self_adj2} and \eqref{eq:ad_self_adj3} is based on Green's first identity.

Finally, due to linearity, we can combine both operators and conclude that $P^{AD}$ is not self-adjoint (nor anti-self-adjoint).

\begin{sloppypar}
\begin{remark}\label{rem:gh_nsa}
Note that based on a similar derivation, it can be shown that the limit operators of $G_\epsilon$ and $H_\epsilon$, i.e. $G=\lim_{\epsilon\rightarrow 0}\left(G_\epsilon-I\right)/\epsilon^2$ and $H=\lim_{\epsilon\rightarrow 0}\left(H_\epsilon-I\right)/\epsilon^2$ , are not self-adjoint as well.
\end{remark}
\begin{remark}\label{rem:diff_ad_forth}
When reversing the kernel order, i.e. $\tilde{P}^{AD}_\epsilon f(x)=P_\epsilon^{(1)}\phi^*P_\epsilon^{(2)}(\phi^*)^{-1}f(x)$, the asymptotic expansion of the resulting alternating diffusion operator is given by a similar expression, up to the forth order terms, $O(\epsilon^4)$. Therefore, constructing the difference operator, $A_\epsilon$ from Subsection \ref{sub:theoNew}, using two alternating diffusion operators with reversed order, i.e. $A_\epsilon^{AD}f(x) = \frac{1}{2}(P^{AD}_\epsilon - \tilde{P}^{AD}_\epsilon )f(x)$, will result in cancellation of all second order terms, $A_\epsilon^{AD}f(x)=O(\epsilon^4)$.
\end{remark}
\end{sloppypar}

\section{Proof of Proposition \ref{prop:SA_sa}\label{app:propSA_sa}}

\begin{sloppypar}
Define the limit operator of $A_{\epsilon_1,\epsilon_2}$, where $\epsilon_2=\alpha\epsilon$ and $\epsilon_1=\epsilon$, $\alpha>0$, by $A_{\alpha} = \lim_{\epsilon\rightarrow 0} A_{\epsilon_1,\epsilon_2}/\epsilon^2$. We show in this appendix that $jA_{\alpha}$ is self-adjoint, by equivalently showing that $A_{\alpha}$ is anti-self-adjoint.
\end{sloppypar}

The asymptotic expansion of $A_{\alpha}: C^{\infty}(\mathcal{M}^{(1)}) \rightarrow C^{\infty}(\mathcal{M}^{(1)})$ is given by:
\begin{eqnarray}
A_{\alpha}f(x) &= & \frac{1}{2}\left(\frac{2\nabla^{(1)}f\cdot\nabla^{(1)}\mu^{(1)}}{\mu^{(1)}}(x) + \frac{f\Delta^{(1)}\mu^{(1)}}{\mu^{(1)}}(x)\right)\label{eq:a_exp_m1}\\
& & - \frac{\alpha^2}{2}\left(\phi^*\frac{2\nabla^{(2)}(\phi^*)^{-1}f\cdot\nabla^{(2)}\mu^{(2)}}{\mu^{(2)}}(x) + f\phi^*\frac{\Delta^{(2)}\mu^{(2)}} {\mu^{(2)}}(x)\right).\label{eq:a_exp_m2}
\end{eqnarray}
This is obtained from Proposition \ref{prop:SA}, for $A_{\epsilon_1,\epsilon_2}/\epsilon^2$ when $\epsilon\rightarrow 0$ and $\epsilon_2=\alpha\epsilon_1=\alpha\epsilon$.

\begin{proof}
Denote by $A^{(1)}_{\alpha}$ the terms in the asymptotic expansion of $A_{\alpha}$ which are related to the first manifold, i.e. \eqref{eq:a_exp_m1}. 
Similarly, denote by $A^{(2)}_{\alpha}$ the terms which are related to the second manifold, i.e. \eqref{eq:a_exp_m2}. In order to show that $A_{\alpha}$ is anti-self-adjoint we will first show that each of these partial operators are anti-self-adjoint and then, from the linearity of the inner product and the additivity of these terms, this result naturally extends to $A_{\alpha}$.

For $A^{(1)}_{\alpha}$ , given $f,g\in C^{\infty}(\mathcal{M}^{(1)})$, 
\begin{align}
\left\langle A^{(1)}_{\alpha}f,g \right\rangle_{\mathcal{M}^{(1)}} & = \int_{\mathcal{M}^{(1)}}\left(\frac{f\Delta^{(1)}\mu^{(1)}}{2\mu^{(1)}} + \frac{\nabla^{(1)}f\cdot\nabla^{(1)}\mu^{(1)}}{\mu^{(1)}}\right)(x)g(x)\mu^{(1)}(x)dV^{(1)}(x)\\
& = \int_{\mathcal{M}^{(1)}}\left(\frac{1}{2}f\Delta^{(1)}\mu^{(1)}\right)(x)g(x)dV^{(1)}(x)\\ 
& + \int_{\mathcal{M}^{(1)}}\left(\nabla^{(1)}f\cdot\nabla^{(1)}\mu^{(1)}\right)(x)g(x)dV^{(1)}(x)\label{eq:a1_self_adj_preGreen}\\
& = \int_{\mathcal{M}^{(1)}}\left(\frac{1}{2}fg\Delta^{(1)}\mu^{(1)}\right)(x)dV^{(1)}(x)\\
& -\int_{\mathcal{M}^{(1)}}\left(\nabla^{(1)}\cdot\left(g\nabla^{(1)}\mu^{(1)}\right)\right)(x)f(x)dV^{(1)}(x)\label{eq:a1_self_adj_postGreen}\\
& = -\int_{\mathcal{M}^{(1)}}\left(\frac{1}{2}g\Delta^{(1)}\mu^{(1)} + \nabla^{(1)}g\nabla^{(1)}\mu^{(1)}\right)(x)f(x)dV^{(1)}(x)\\
& = \int_{\mathcal{M}^{(1)}}\left(\frac{g\Delta^{(1)}\mu^{(1)}}{2\mu^{(1)}} + \frac{\nabla^{(1)}g\cdot\nabla^{(1)}\mu^{(1)}}{\mu^{(1)}}\right)(x)f(x)\mu^{(1)}(x)dV^{(1)}(x)\\
 & = -\left\langle f,A^{(1)}_{\alpha}g \right\rangle_{\mathcal{M}^{(1)}},
\end{align}
where the transition between \eqref{eq:a1_self_adj_preGreen} and \eqref{eq:a1_self_adj_postGreen} is based on Green's first identity (for manifolds without a boundary).

Similarly, for $A^{(2)}_{\alpha}$, given $f,g\in C^{\infty}(\mathcal{M}^{(1)})$, 
\begin{align}
\left\langle A^{(2)}_{\alpha}f,g \right\rangle_{\mathcal{M}^{(1)}} & = -\int_{\mathcal{M}^{(1)}}\alpha^2\left(f\phi^*\frac{\Delta^{(2)}\mu^{(2)}} {2\mu^{(2)}}\right)(x)g(x)\mu^{(1)}(x)dV^{(1)}(x)\nonumber\\
& -\int_{\mathcal{M}^{(1)}}\alpha^2\left(\phi^*\frac{\nabla^{(2)}(\phi^*)^{-1}f\cdot\nabla^{(2)}\mu^{(2)}}{\mu^{(2)}}\right)(x)g(x)\mu^{(1)}(x)dV^{(1)}(x)\label{eq:a2_self_adj1}\\
& = -\int_{\mathcal{M}^{(2)}}\alpha^2\left((\phi^*)^{-1}f\frac{\Delta^{(2)}\mu^{(2)}} {2\mu^{(2)}}(\phi^*)^{-1}g\right)(y)\mu^{(2)}(y)dV^{(2)}(y) \nonumber\\
& -\int_{\mathcal{M}^{(2)}}\alpha^2\left(\frac{\nabla^{(2)}(\phi^*)^{-1}f\cdot\nabla^{(2)}\mu^{(2)}}{\mu^{(2)}}(\phi^*)^{-1}g\right)(y)\mu^{(2)}(y)dV^{(2)}(y)\label{eq:a2_self_adj2}\\
& = -\int_{\mathcal{M}^{(2)}}\alpha^2\left(\frac{1}{2}(\phi^*)^{-1}g(\phi^*)^{-1}f\Delta^{(2)}\mu^{(2)}\right)(y)dV^{(2)}(y)\nonumber\\ 
& + \int_{\mathcal{M}^{(2)}}\alpha^2\left((\phi^*)^{-1}g\Delta^{(2)}\mu^{(2)}(\phi^*)^{-1}f\right)(y)dV^{(2)}(y) \nonumber\\
& + \int_{\mathcal{M}^{(2)}}\alpha^2\left(\nabla^{(2)}(\phi^*)^{-1}g\cdot\nabla^{(2)}\mu^{(2)}(\phi^*)^{-1}f\right)(y)dV^{(2)}(y)\label{eq:a2_self_adj3}\\
& = \int_{\mathcal{M}^{(2)}}\alpha^2\left((\phi^*)^{-1}g\frac{\Delta^{(2)}\mu^{(2)}}{2\mu^{(2)}}(\phi^*)^{-1}f\right)(y)\mu^{(2)}(y)dV^{(2)}(y) \nonumber\\
& + \int_{\mathcal{M}^{(2)}}\alpha^2\left(\frac{\nabla^{(2)}(\phi^*)^{-1}g\cdot\nabla^{(2)}\mu^{(2)}}{\mu^{(2)}}(\phi^*)^{-1}f\right)(y)\mu^{(2)}(y)dV^{(2)}(y)\label{eq:a2_self_adj4}\\
& = \int_{\mathcal{M}^{(1)}}\alpha^2\left(g\phi^*\frac{\Delta^{(2)}\mu^{(2)}}{2\mu^{(2)}}\right)(x)f(x)\mu^{(1)}(x)dV^{(1)}(x)\nonumber\\
& + \int_{\mathcal{M}^{(1)}}\alpha^2\left(\phi^*\frac{\nabla^{(2)}(\phi^*)^{-1}g\cdot\nabla^{(2)}\mu^{(2)}}{\mu^{(2)}}\right)(x)f(x)\mu^{(1)}(x)dV^{(1)}(x)\label{eq:a2_self_adj5}\\
 & = -\left\langle f,A^{(2)}_{\alpha}g \right\rangle_{\mathcal{M}^{(1)}},
\end{align}
where the transitions from \eqref{eq:a2_self_adj1} to \eqref{eq:a2_self_adj2} and from \eqref{eq:a2_self_adj4} to \eqref{eq:a2_self_adj5} are based on $\mu^{(1)}(x)dV^{(1)}(x)=\mu^{(2)}(y)dV^{(2)}(y)$ and $y=\phi(x)$. In addition, the transition between \eqref{eq:a2_self_adj2} and \eqref{eq:a2_self_adj3} is based on Green's first identity.

Finally, combining these results for $A^{(1)}_{\alpha}$ and $A^{(2)}_{\alpha}$ we get: 
\begin{eqnarray}
\left\langle jA_{\alpha}f,g\right\rangle_{\mathcal{M}^{(1)}} & = & \left\langle j\left(A^{(1)}_{\alpha}+A^{(2)}_{\alpha}\right)f,g\right\rangle_{\mathcal{M}^{(1)}}\\
& = & j\left\langle A^{(1)}_{\alpha}f,g\right\rangle_{\mathcal{M}^{(1)}} + j\left\langle A^{(2)}_{\alpha}f,g\right\rangle_{\mathcal{M}^{(1)}}\\
& = & -j\left\langle f,-A^{(1)}_{\alpha}g\right\rangle_{\mathcal{M}^{(1)}} - j\left\langle f,A^{(2)}_{\alpha}g\right\rangle_{\mathcal{M}^{(1)}}\\
& = & -j\left\langle f,\left(A^{(1)}_{\alpha}+A^{(2)}_{\alpha}\right)g\right\rangle_{\mathcal{M}^{(1)}} = \left\langle f,jA_{\alpha}g\right\rangle_{\mathcal{M}^{(1)}}.
\end{eqnarray}
\end{proof}

\begin{remark}
By performing a similar derivation for the operator $S_\epsilon$, it can be shown to be self-adjoint as well.
\end{remark}

\section{Proof of Proposition \ref{prop:Asupp}}\label{app:propAsupp}

We prove here that $\forall f\in C^{\infty}\left(\mathcal{M}^{(1)}\right)$, if $\mathrm{supp}f\subset\mathring{\Omega}_{\alpha}$, then $A_{\alpha}f(x)=0$, where, as defined in Section \ref{sec:formulation}, $\Omega_{\alpha}=\left\lbrace x\in\mathcal{M}^{(1)}: \ \nabla\phi\vert_x=\alpha\mathrm{I}\right\rbrace$, $\alpha>0$.

\begin{proof}
As presented in Proposition \ref{prop:SA} and in Appendix \ref{app:propSA_sa}, the asymptotic expansion of the operator $A_{\alpha}$ is given by
\begin{eqnarray}
A_{\alpha}f(x) = & & \frac{1}{2}\left(\frac{2\nabla^{(1)}f\cdot\nabla^{(1)}\mu^{(1)}}{\mu^{(1)}}(x) + \frac{f\Delta^{(1)}\mu^{(1)}}{\mu^{(1)}}(x)\right)\\
& & - \frac{\alpha^2}{2}\left(\phi^*\frac{2\nabla^{(2)}(\phi^*)^{-1}f\cdot\nabla^{(2)}\mu^{(2)}}{\mu^{(2)}}(x) + f\phi^*\frac{\Delta^{(2)}\mu^{(2)}} {\mu^{(2)}}(x)\right).\label{eq:A_appF}
\end{eqnarray}

Consider $x\in\mathcal{M}^{(1)}$, $y=\phi(x)\in\mathcal{M}^{(2)}$ and $f\in C^{\infty}\left(\mathcal{M}^{(1)}\right)$. With the chosen coordinates around $x$ and $y$, we calculate the following gradient of $f$:
\begin{equation}
\left.\left(\nabla^{(2)}(\phi^*)^{-1}f\right)\right\vert_y=\left.\left(\nabla^{(2)}f\circ\phi^{-1}\right)\right\vert_y=\nabla^{(1)}f\vert_x\nabla^{(2)}\phi^{-1}\vert_y.
\end{equation}
In addition, calculating the gradient of the density function of the manifold $\mathcal{M}^{(2)}$, given by $\mu^{(2)}(y)=J(y)\mu^{(1)}\left(\phi^{-1}(y)\right)$, where $J(y)=\left\vert det\left(\nabla^{(2)}\phi^{-1}(y)\right)\right\vert$, leads to:
\begin{eqnarray}
\nabla^{(2)}\mu^{(2)}\vert_y & = & \nabla^{(2)}\left.\left(J\mu^{(1)}\circ\phi^{-1}\right)\right\vert_y\\
& = & \nabla^{(2)}J\vert_y\left.\left(\mu^{(1)}\circ\phi^{-1}\right)\right\vert_y + J\vert_y\nabla^{(1)}\mu^{(1)}\vert_x\nabla^{(2)}\phi^{-1}\vert_y.
\end{eqnarray}

By substituting these derivations in expression \eqref{eq:A_appF}, we get:
\begin{eqnarray}
A_{\alpha}f(x) = & & \frac{1}{2}\left(\frac{2\nabla^{(1)}f\cdot\nabla^{(1)}\mu^{(1)}}{\mu^{(1)}}(x) + \frac{f\Delta^{(1)}\mu^{(1)}}{\mu^{(1)}}(x)\right)\\
& & - \frac{\alpha^2}{2} \frac{2\nabla^{(1)}f\vert_x\nabla^{(2)}\phi^{-1}\vert_{\phi(x)}\cdot\nabla^{(2)}J\vert_{\phi(x)}\mu^{(1)}}{J\vert_{\phi(x)}\mu^{(1)}\vert_x}\\
& & -\frac{\alpha^2}{2}\frac{2\nabla^{(1)}f\vert_x\nabla^{(2)}\phi^{-1}\vert_{\phi(x)}\cdot \nabla^{(1)}\mu^{(1)}\vert_x\nabla^{(2)}\phi^{-1}\vert_{\phi(x)}}{\mu^{(1)}\vert_x}\\
& & -\frac{\alpha^2}{2}f\frac{\Delta^{(2)}\mu^{(2)}\vert_{\phi(x)}} {\mu^{(2)}\vert_{\phi(x)}}.
\end{eqnarray}

Then, if $\mathrm{supp}f\subset\mathring{\Omega}_{\alpha}$, for $x\in\mathring{\Omega}_{\alpha}$ we have $\nabla^{(2)}\phi^{-1}\vert_{\phi(x)}=\frac{1}{\alpha}\mathrm{I}$, where $\mathrm{I}$ denotes the $d\times d$ identity matrix, and $J\vert_{\phi(x)}=\alpha^{-d}$. In addition, for such $x$, we have $\mu^{(2)}(\phi(x))=\alpha^{-d}\mu^{(1)}(x)$.
We are then left with:
\begin{eqnarray}
A_{\alpha}f(x) & = & \frac{1}{2}\left(\frac{2\nabla^{(1)}f\cdot\nabla^{(1)}\mu^{(1)}}{\mu^{(1)}}(x) + \frac{f\Delta^{(1)}\mu^{(1)}}{\mu^{(1)}}(x)\right)\\
& & -\frac{\alpha^2}{2}\left(\frac{2\nabla^{(1)}f\alpha^{-1}\cdot\nabla^{(1)}\mu^{(1)}\alpha^{-1}}{\mu^{(1)}}(x)
+f\frac{\Delta^{(2)}\mu^{(2)}}{\mu^{(2)}}(\phi(x))\right)\\
& = & \frac{1}{2}\left(\frac{f\Delta^{(1)}\mu^{(1)}}{\mu^{(1)}}(x) - \alpha^2 f\frac{\Delta^{(2)}\mu^{(2)}} {\mu^{(2)}}(\phi(x))\right)\\
& = & \frac{1}{2}\left(\frac{f\Delta^{(1)}\mu^{(1)}}{\mu^{(1)}}(x) - \alpha^2f\frac{\alpha^{-d-2}\Delta^{(1)}\mu^{(1)}} {\alpha^{-d}\mu^{(1)}}(x)\right)\\
& = & 0.
\end{eqnarray}
where we use the fact that for $x\in\mathring{\Omega}_{\alpha}$, $\Delta^{(2)}\mu^{(2)}\left(\phi(x)\right)=\alpha^{-d-2}\Delta^{(1)}\mu^{(1)}(x)$.

Therefore, we showed that if $\mathrm{supp}f\subset\mathring{\Omega}_{\alpha}$, then $A_{\alpha}f(x)=0$.
\end{proof}

\section{Proof of Corollary \ref{cor:Asupp}}
In this appendix we prove Corollary \ref{cor:Asupp}. For simplicity, we assume here that $\epsilon_2=\epsilon_1=\epsilon$ ($\alpha=1$). For $\epsilon_2\neq\epsilon_1$, the derivations are similar up to some notation changes, as in Appendix \ref{app:propAsupp}.

Consider $\mathcal{E}^{(1)} \subset \mathbb{R}^p$ and $\mathcal{E}^{(2)} \subset \mathbb{R}^p$ such that $\mathcal{E}^{(\ell)} = \mathcal{M}^{(\ell)} \oplus \mathcal{F}^{(\ell)}$, where $\mathcal{M}^{(\ell)}\subset\mathbb{R}^{p_1}$, $\mathcal{F}^{(\ell)}\subset\mathbb{R}^{p_2}$, $p=p_1+p_2$, $\ell=1,2$, and $\phi:\mathcal{E}^{(1)}\rightarrow\mathcal{E}^{(2)}$ satisfies $\phi(\mathcal{M}^{(1)} \oplus \mathcal{F}^{(1)})=\mathcal{M}^{(1)} \oplus \tilde{\phi}(\mathcal{F}^{(1)})$, where $\tilde{\phi}:\mathcal{F}^{(1)}\rightarrow\mathcal{F}^{(2)}$ is a smooth diffeomorphism. 
In addition, assume that $\mu^{(\ell)}(\mathbf{s}^{(\ell)})=\mu^{(\ell)}_m(\mathbf{m}^{(\ell)})\mu^{(\ell)}_f(\mathbf{f}^{(\ell)})$, where $\mu^{(\ell)}$ is the probability density on $\mathcal{E}^{(\ell)}$, $\mu^{(\ell)}_m$ is the marginal density of $\mu^{(\ell)}$ on $\mathcal{M}^{(\ell)}$, $\mu^{(\ell)}_f$ is the marginal density of $\mu^{(\ell)}$ on $\mathcal{F}^{(\ell)}$ and $\mathbf{s}^{(\ell)}(t)=\mathbf{m}^{(\ell)}(t)+\mathbf{f}^{(\ell)}(t)$, where $\mathbf{s}^{(\ell)}\in\mathcal{E}^{(\ell)}$, $\mathbf{m}^{(\ell)}\in\mathcal{M}^{(\ell)}$ and $\mathbf{f}^{(\ell)}\in\mathcal{F}^{(\ell)}$.

Denote $\Omega_f = \left\{\mathbf{f}^{(1)}(t) \in \mathcal{F}^{(1)}: \ \nabla \tilde{\phi} | _{\mathbf{f}^{(1)}} = \mathrm{I}\right\}\subset\mathcal{F}^{(1)}$, where $\mathrm{I}$ denotes a $p_2\times p_2$ identity matrix, and define $A = \lim_{\epsilon\rightarrow 0} A_\epsilon/\epsilon^2$.

Corollary \ref{cor:Asupp} states that for all $g\in C^\infty\left(\mathcal{E}^{(1)}\right)$, if $\textmd{supp}g \subset \mathcal{M}^{(1)} \oplus \mathring{\Omega}_f$, then $A g=0$.
Hence, if $A g = \lambda g$, $g \neq 0$, then, $\textmd{supp}g \subset \mathcal{M}^{(1)} \oplus \Omega_f^c$.

\begin{proof}
We first note that since $\mathcal{E}^{(1)}=\mathcal{M}^{(1)}\oplus\mathcal{F}^{(1)}$, the eigenfunctions of $A\vert_{\mathcal{F}^{(1)}}$, i.e. the restriction of $A$ to $\mathcal{F}^{(1)}$, multiplied by a non-zero function defined on $\mathcal{M}^{(1)}$, are eigenfunctions of $A$.
Second, note that $\nabla^{(1)}\phi\neq\mathrm{I}$ when $\nabla^{(1)}\tilde{\phi}\neq\mathrm{I}$, since
\begin{equation}
\nabla^{(1)}\phi_{p\times p} = 
\begin{bmatrix}
\mathrm{I}_{p_1\times p_1} & \boldsymbol{0}_{p_1\times p_2} \\
\boldsymbol{0}_{p_2\times p_1} & \nabla^{(1)}\tilde{\phi}_{p_2\times p_2}
\end{bmatrix}
\end{equation}
where $\boldsymbol{0}_{d_1\times d_2}$ denotes a zero matrix of size $d_1\times d_2$.
Third, from the relation between the probability density functions on the two manifolds, we have $\mu^{(2)}_m(\mathbf{m}^{(2)})=\mu^{(1)}_m(\mathbf{m}^{(1)})$ and $\mu^{(2)}_f(\mathbf{f}^{(2)})=\left.J_{\tilde{\phi}}\right\vert_{\mathbf{f}^{(2)}}\mu^{(1)}_f(\mathbf{f}^{(1)})$, where $\left.J_{\tilde{\phi}}\right\vert_{\mathbf{f}^{(2)}}=\left\vert \mathrm{det}\left(\nabla^{(2)}\tilde{\phi}^{-1}(\mathbf{f}^{(2)})\right)\right\vert$, since $\left.J_{\phi}\right\vert_{\mathbf{s}^{(2)}}=\left.J_{\tilde{\phi}}\right\vert_{\mathbf{f}^{(2)}}$ and $\mu^{(2)}(\mathbf{s}^{(2)})=\left.J_{\phi}\right\vert_{\mathbf{s}^{(2)}}\mu^{(1)}(\mathbf{s}^{(1)})$, $\mathbf{s}^{(2)}=\phi(\mathbf{s}^{(1)})$.

Therefore, we can derive the following expressions for $g\in C^{\infty}\left(\mathcal{E}^{(1)}\right)$, $\phi^{-1}$ and $\mu^{(\ell)}$:
\begin{eqnarray}
\nabla^{(1)}g\vert_{\mathbf{s}^{(1)}} = 
\begin{bmatrix}
\left.\nabla^{(1)}_m g\right\vert_{\mathbf{m}^{(1)}}\\
\left.\nabla^{(1)}_f g\right\vert_{\mathbf{f}^{(1)}}
\end{bmatrix}
& \left.\nabla^{(2)}\phi^{-1}\right\vert_{\phi(\mathbf{s}^{(1)})}=
\begin{bmatrix}
\mathrm{I}_{p_1\times p_1}     & \boldsymbol{0}_{p_1\times p_2}\\
\boldsymbol{0}_{p_2\times p_1} & \left.\nabla^{(2)}\tilde{\phi}^{-1}\right\vert_{\phi(\mathbf{f}^{(1)})}
\end{bmatrix}\label{eq:appg_deriv1}
\end{eqnarray}
\begin{eqnarray}
\left.\nabla^{(\ell)}\mu^{(\ell)}\right\vert_{\mathbf{s}^{(1)}} = 
\begin{bmatrix}
\mu^{(\ell)}_f(\mathbf{f}^{(\ell)})\left.\nabla^{(\ell)}_m\mu^{(\ell)}_m\right\vert_{\mathbf{m}^{(\ell)}}\\
\left.\nabla^{(\ell)}_f\mu^{(\ell)}_f\right\vert_{\mathbf{f}^{(\ell)}}\mu^{(\ell)}_m(\mathbf{m}^{(\ell)})\\
\end{bmatrix} 
\end{eqnarray}
\begin{eqnarray}
\left.\Delta^{(\ell)}\mu^{(\ell)}\right\vert_{\mathbf{s}^{(\ell)}} =
\mu^{(\ell)}_f(\mathbf{f}^{(\ell)})\left.\Delta^{(\ell)}_m\mu^{(\ell)}_m\right\vert_{\mathbf{m}^{(\ell)}} + 
\left.\Delta^{(\ell)}_f\mu^{(\ell)}_f\right\vert_{\mathbf{f}^{(\ell)}}\mu^{(\ell)}_m(\mathbf{m}^{(\ell)})
\label{eq:appg_deriv2}
\end{eqnarray}
\begin{eqnarray}
\nabla^{(1)}g\vert_{\mathbf{s}^{(1)}}\nabla^{(2)}\phi^{-1}\vert_{\phi(\mathbf{s}^{(1)})} = &
\begin{bmatrix}
\nabla^{(1)}_m g\vert_{\mathbf{m}^{(1)}}\\
\nabla^{(1)}_f g\vert_{\mathbf{f}^{(1)}}\left.\nabla^{(2)}\tilde{\phi}^{-1}\right\vert_{\tilde{\phi}(\mathbf{f}^{(1)})}
\end{bmatrix}\label{eq:appg_deriv3}
\\
\nabla^{(1)}\mu^{(1)}\vert_{\mathbf{s}^{(1)}}\nabla^{(2)}\phi^{-1}\vert_{\phi(\mathbf{s}^{(1)})} = &
\begin{bmatrix}
\mu^{(1)}_f(\mathbf{f}^{(1)})\left.\nabla^{(1)}_m\mu^{(1)}_m\right\vert_{\mathbf{m}^{(1)}}\\
\left.\nabla^{(1)}_f\mu^{(1)}_f\right\vert_{\mathbf{f}^{(1)}}\left.\nabla^{(2)}\tilde{\phi}^{-1}\right\vert_{\tilde{\phi}(\mathbf{f}^{(1)})}\mu^{(1)}_m(\mathbf{m}^{(1)})
\end{bmatrix}.\label{eq:appg_deriv4}
\end{eqnarray}

According to Appendix \ref{app:propAsupp} the operator $A = \lim_{\epsilon\rightarrow 0} A_\epsilon/\epsilon^2$ is given by
\begin{eqnarray}
A g(x) = & & \frac{1}{2}\left(\frac{2\nabla^{(1)}g\cdot\nabla^{(1)}\mu^{(1)}}{\mu^{(1)}}(x) + \frac{g\Delta^{(1)}\mu^{(1)}}{\mu^{(1)}}(x)\right)\nonumber\\
& & - \frac{1}{2} \frac{2\nabla^{(1)}g\vert_x\nabla^{(2)}\phi^{-1}\vert_{\phi(x)}\cdot\nabla^{(2)}J\vert_{\phi(x)}\mu^{(1)}}{J\vert_{\phi(x)}\mu^{(1)}\vert_x}\nonumber\\
& & -\frac{1}{2}\frac{2\nabla^{(1)}g\vert_x\nabla^{(2)}\phi^{-1}\vert_{\phi(x)}\cdot \nabla^{(1)}\mu^{(1)}\vert_x\nabla^{(2)}\phi^{-1}\vert_{\phi(x)}}{\mu^{(1)}\vert_x}\nonumber\\
& & -\frac{1}{2}g\frac{\Delta^{(2)}\mu^{(2)}\vert_{\phi(x)}} {\mu^{(2)}\vert_{\phi(x)}}.\label{eq:appg_A}
\end{eqnarray}
By substituting expressions \eqref{eq:appg_deriv1} - \eqref{eq:appg_deriv4} and $\mu^{(\ell)}(\mathbf{s}^{(1)})=\mu^{(\ell)}_m(\mathbf{m}^{(1)})\mu^{(\ell)}_f(\mathbf{f}^{(1)})$ into \eqref{eq:appg_A}, we get:
\begin{eqnarray}
A g(\mathbf{s}^{(1)}) = & & \frac{1}{2}\left(\frac{2\nabla^{(1)}_m g\vert_{\mathbf{m}^{(1)}}\cdot\nabla^{(1)}_m\mu^{(1)}_m\vert_{\mathbf{m}^{(1)}}}{\mu^{(1)}_m\vert_{\mathbf{m}^{(1)}}}
+ \frac{g\Delta^{(1)}_m\mu^{(1)}_m\vert_{\mathbf{m}^{(1)}}}{\mu^{(1)}_m\vert_{\mathbf{m}^{(1)}}}\right)\nonumber\\
& & + \frac{1}{2}\left(\frac{2\nabla^{(1)}_f g\vert_{\mathbf{f}^{(1)}}\cdot\nabla^{(1)}_f\mu^{(1)}_f\vert_{\mathbf{f}^{(1)}}}{\mu^{(1)}_f\vert_{\mathbf{f}^{(1)}}}
+ \frac{g\Delta^{(1)}_f\mu^{(1)}_f\vert_{\mathbf{f}^{(1)}}}{\mu^{(1)}_f\vert_{\mathbf{f}^{(1)}}}\right)\nonumber\\
& & - \frac{1}{2} \frac{2\nabla^{(1)}_f g\vert_{\mathbf{f}^{(1)}}\nabla^{(2)}\tilde{\phi}^{-1}\vert_{\tilde{\phi}(\mathbf{f}^{(1)})}\cdot\nabla^{(2)}J_{\tilde{\phi}}\vert_{\tilde{\phi}(\mathbf{f}^{(1)})}\mu^{(1)}_f\vert_{\mathbf{f}^{(1)}}}{J_{\tilde{\phi}}\vert_{\tilde{\phi}(\mathbf{f}^{(1)})}\mu^{(1)}_f\vert_{\mathbf{f}^{(1)}}}\nonumber\\
& & -\frac{1}{2}\frac{2\nabla^{(1)}_m g\vert_{\mathbf{m}^{(1)}}\cdot\nabla^{(1)}_m\mu^{(1)}_m\vert_{\mathbf{m}^{(1)}}}{\mu^{(1)}_m\vert_{\mathbf{m}^{(1)}}}\nonumber\\
& & -\frac{1}{2}\frac{2\nabla^{(1)}_f g\vert_{\mathbf{f}^{(1)}}\nabla^{(2)}\tilde{\phi}^{-1}\vert_{\tilde{\phi}(\mathbf{f}^{(1)})}\cdot \nabla^{(1)}_f\mu^{(1)}_f\vert_{\mathbf{f}^{(1)}}\nabla^{(2)}\tilde{\phi}^{-1}\vert_{\tilde{\phi}(\mathbf{f}^{(1)})}}{\mu^{(1)}_f\vert_{\mathbf{f}^{(1)}}}\nonumber\\
& & -\frac{1}{2}\frac{g\Delta^{(2)}_f\mu^{(2)}_f\vert_{\tilde{\phi}(\mathbf{f}^{(1)})}} {\mu^{(2)}_f\vert_{\tilde{\phi}(\mathbf{f}^{(1)})}}
-\frac{1}{2}\frac{g\Delta^{(1)}_m\mu^{(1)}_m\vert_{\mathbf{m}^{(1)}}} {\mu^{(1)}_m\vert_{\mathbf{m}^{(1)}}}\label{eq:appg_finalA1}\\
= & & \frac{1}{2}\left(\frac{2\nabla^{(1)}_f g\vert_{\mathbf{f}^{(1)}}\cdot\nabla^{(1)}_f\mu^{(1)}_f\vert_{\mathbf{f}^{(1)}}}{\mu^{(1)}_f\vert_{\mathbf{f}^{(1)}}}
+ \frac{g\Delta^{(1)}_f\mu^{(1)}_f\vert_{\mathbf{f}^{(1)}}}{\mu^{(1)}_f\vert_{\mathbf{f}^{(1)}}}\right)\nonumber\\
& & - \frac{1}{2} \frac{2\nabla^{(1)}_f g\vert_{\mathbf{f}^{(1)}}\nabla^{(2)}\tilde{\phi}^{-1}\vert_{\tilde{\phi}(\mathbf{f}^{(1)})}\cdot\nabla^{(2)}J_{\tilde{\phi}}\vert_{\tilde{\phi}(\mathbf{f}^{(1)})}\mu^{(1)}_f\vert_{\mathbf{f}^{(1)}}}{J_{\tilde{\phi}}\vert_{\tilde{\phi}(\mathbf{f}^{(1)})}\mu^{(1)}_f\vert_{\mathbf{f}^{(1)}}}\nonumber\\
& & -\frac{1}{2}\frac{2\nabla^{(1)}_f g\vert_{\mathbf{f}^{(1)}}\nabla^{(2)}\tilde{\phi}^{-1}\vert_{\tilde{\phi}(\mathbf{f}^{(1)})}\cdot \nabla^{(1)}_f\mu^{(1)}_f\vert_{\mathbf{f}^{(1)}}\nabla^{(2)}\tilde{\phi}^{-1}\vert_{\tilde{\phi}(\mathbf{f}^{(1)})}}{\mu^{(1)}_f\vert_{\mathbf{f}^{(1)}}}\nonumber\\
& & -\frac{1}{2}\frac{g\Delta^{(2)}_f\mu^{(2)}_f\vert_{\tilde{\phi}(\mathbf{f}^{(1)})}}{\mu^{(2)}_f\vert_{\tilde{\phi}(\mathbf{f}^{(1)})}}\\
= & & A\vert_{\mathcal{F}^{(1)}}g(\mathbf{f}^{(1)}).\label{eq:appg_finalA2}
\end{eqnarray}
where we used $\mu^{(2)}_m\vert_{\phi(\mathbf{m}^{(1)})}=\mu^{(1)}_m\vert_{\mathbf{m}^{(1)}}$ and $\Delta^{(2)}_m\mu^{(2)}_m\vert_{\phi(\mathbf{m}^{(1)})}=\Delta^{(1)}_m\mu^{(1)}_m\vert_{\mathbf{m}^{(1)}}$ to obtain the last term in \eqref{eq:appg_finalA1}.

This derivation states that $Ag(\mathbf{s}^{(1)})=A\vert_{\mathcal{F}^{(1)}}g(\mathbf{f}^{(1)})$. Therefore, under the assumptions stated in the beginning of this appendix, the considered setting is equivalent to the setting in Proposition \ref{prop:Asupp}, with the manifolds $\mathcal{F}^{(\ell)}$, $\ell=1,2$, the smooth diffeomorphism $\tilde{\phi}:\mathcal{F}^{(1)}\rightarrow\mathcal{F}^{(2)}$ and $g\in C^{\infty}\left(\mathcal{F}^{(1)}\right)$. We can now apply Proposition \ref{prop:Asupp} to \eqref{eq:appg_finalA2} and obtain that for all $g\in C^{\infty}\left(\mathcal{F}^{(1)}\right)$, if $\textmd{supp}g\subset\mathring{\Omega_f}$, then $A\vert_{\mathcal{F}^{(1)}}g(\mathbf{f}^{(1)})=0$. 
Due to the definition of $\mathcal{E}^{(\ell)}$ as a direct sum of $\mathcal{M}^{(\ell)}$ and $\mathcal{F}^{(\ell)}$, we can define $g\in C^{\infty}\left(\mathcal{E}^{(1)}\right)$ and obtain that for all $g\in C^{\infty}\left(\mathcal{E}^{(1)}\right)$, if $\textmd{supp}g \subset \mathcal{M}^{(1)} \oplus \mathring{\Omega}_f$, then $Ag(\mathbf{s}^{(1)})=0$, which concludes the proof.
\end{proof}

\end{appendices}

\bibliographystyle{siamplain}
\bibliography{papersNew}

\end{document}